\newif\ifShowKeys
\ShowKeystrue
\ShowKeysfalse

\documentclass[letterpaper,10pt]{article}

\pdfoutput=1
\usepackage[no-natbib-sort]{my-jheppub}

\usepackage{amsmath, amssymb}

\usepackage{xfrac}   

\usepackage{amsthm}

\usepackage{bm,environ,mathrsfs,array,arydshln,booktabs,float,slashed,hyperref}
\usepackage[mathcal]{euscript}
\usepackage{tensor} 	
\usepackage{esint}

\usepackage[nodayofweek]{date time}

\font\sm=cmss10 at 4pt
\usepackage{graphicx,epsfig}
\usepackage{epic}
\usepackage{tikz} 
\usetikzlibrary{decorations.pathreplacing,decorations.markings,snakes}

\tikzset{middlearrow/.style={
        decoration={markings,
            mark= at position 0.5 with {\arrow{#1}} ,
        },
        postaction={decorate}
    }
}

\usepackage{aurical}
\usepackage[T1]{fontenc}

\usepackage{framed}
\definecolor{shadecolor}{RGB}{255, 230, 204}

\allowdisplaybreaks

\newcommand{\be}{\begin{equation}}
\newcommand{\ee}{\end{equation}}
\newcommand{\mc}{\mathcal }

\newcommand{\la}{\label}
\newcommand{\eps}{\varepsilon}
\newcommand{\pb}{\overline{\partial}}

\newcommand{\wb}{\overline{w}}
\newcommand{\psib}{\tilde{\psi}}

\newcommand{\sfPhi}{\mathsf{\Phi}}
\newcommand{\sfPsi}{\mathsf{\Psi}}

\def \bz {\mathsf{z}}
\def \bt {\mathsf{t}}
\def \vp {\varphi}

 \def \OO {{\cal O}} \def \ed {\end{document}}
 \def \iffa {\iffalse} 
\def \ci {\cite} \def \s  {\sigma}
\def\foot{\ \footnote}
\newcommand{\rf}[1]{(\ref{#1})}
\def\la{\label}
\def\l{\lambda}
 \def \te {\textstyle} 
\def \ha {{\te{1\ov 2}}}\def \del  {\partial } 
\def \adst  {{AdS$_2$\ }}
\def \sm {$\s$-model }

\def \k {\kappa}

 \def \no {\nonumber}

\def \sm {$\s$-model\ } \def \sms {$\s$-models\ } 
\def \ov  {\over}

\newcommand{\bpm}{\begin{pmatrix}}
\newcommand{\epm}{\end{pmatrix}}

\newcommand{\PBK}[1]{\ensuremath{\begin{pmatrix}#1\end{pmatrix}}}

\newcommand{\EV}[1]{\langle #1 \rangle}
\newcommand{\beqn}{\begin{eqnarray}}
\newcommand{\eeqn}{\end{eqnarray}}

\usepackage{slashed}

\newcommand{\sfX}{{\sf X}}

\newcommand{\gb}{ g^{_\p}}

\newcommand{\p}{\partial}

\newcommand{\pp}{++}

\newcommand{\rmd}{{\rm d}}

\newcommand{\sft}{{\sf t}}
\newcommand{\sfz}{{\sf z}}

\DeclareMathOperator{\sgn}{sgn}

\DeclareMathOperator{\tr}{tr}

\DeclareMathOperator{\sign}{\text{sign}}

\makeatletter
\DeclareFontFamily{OMX}{MnSymbolE}{}
\DeclareSymbolFont{MnLargeSymbols}{OMX}{MnSymbolE}{m}{n}
\SetSymbolFont{MnLargeSymbols}{bold}{OMX}{MnSymbolE}{b}{n}
\DeclareFontShape{OMX}{MnSymbolE}{m}{n}{
    <-6>  MnSymbolE5
   <6-7>  MnSymbolE6
   <7-8>  MnSymbolE7
   <8-9>  MnSymbolE8
   <9-10> MnSymbolE9
  <10-12> MnSymbolE10
  <12->   MnSymbolE12
}{}
\DeclareFontShape{OMX}{MnSymbolE}{b}{n}{
    <-6>  MnSymbolE-Bold5
   <6-7>  MnSymbolE-Bold6
   <7-8>  MnSymbolE-Bold7
   <8-9>  MnSymbolE-Bold8
   <9-10> MnSymbolE-Bold9
  <10-12> MnSymbolE-Bold10
  <12->   MnSymbolE-Bold12
}{}

\let\llangle\@undefined
\let\rrangle\@undefined
\DeclareMathDelimiter{\llangle}{\mathopen}%
                     {MnLargeSymbols}{'164}{MnLargeSymbols}{'164}
\DeclareMathDelimiter{\rrangle}{\mathclose}%
                     {MnLargeSymbols}{'171}{MnLargeSymbols}{'171}
\makeatother

\title{Boundary correlators in   WZW model on AdS$_2$}
\author[a]{Matteo Beccaria,}
\author[b]{Hongliang Jiang,}
\author[,c]{Arkady A. Tseytlin\footnote{\ Also at the   Institute  for Theoretical and Mathematical Physics, Moscow State University
and  Lebedev Institute, Moscow.}}

\affiliation[a]{Dipartimento di Matematica e Fisica Ennio De Giorgi,\\
Universit\`a del Salento \& INFN, Via Arnesano, 73100 Lecce, 
Italy} 
\affiliation[b]{Albert Einstein Center for Fundamental Physics, Institute for Theoretical Physics, 
University of Bern, \\ Sidlerstrasse 5, 3012 Bern, Switzerland}
\affiliation[c]{Blackett Laboratory, Imperial College, London SW7 2AZ, U.K.}
\emailAdd{matteo.beccaria@le.infn.it,   jiang@itp.unibe.ch,   \\
\qquad\quad \  
 tseytlin@imperial.ac.uk} 

\abstract{Boundary correlators of elementary fields  in  some  2d  conformal field theories defined on  AdS$_2$ 
have a particularly simple structure. For example, the correlators of the Liouville scalar   
happen to be the same as 
the  correlators of  the chiral component of the stress tensor on a plane 
 restricted to the real line. 
Here we show that an analogous relation is true also in the WZW model:  boundary  correlators of the WZW scalars 
 have the same structure as   the correlators of chiral Kac-Moody   currents. 
  This is checked at the level of the tree and  one-loop Witten diagrams in AdS$_2$.
 We  also compute  some  tree-level   correlators in   a generic $\sigma$-model  defined on AdS$_2$
 and show  that they simplify only  in the  WZW  case where an extra Kac-Moody symmetry  appears. 
In particular, the   terms  in 4-point correlators  having    logarithmic dependence on  1d cross-ratio 
cancel only at the WZW  point.  One motivation behind this work is to learn how to compute AdS$_2$
 loop 
corrections  in  2d  models  with derivative interactions  related to the study of 
correlators of operators on Wilson loops  in  string theory in AdS. 
}
\begin{document}


\begin{tabbing}
\hspace*{11.7cm} \=  \kill 
    \> Imperial-TP-AT-2020-01
\end{tabbing}

\maketitle

\def \sG {{\sf  G}}  \def \sX {{\sf X}}
 \def \pp {{\alpha}}
 \def \qq  {{\beta}}
\def \pcmq {PCM$_q$\, }
 \def \sPsi  {\chi} 
 \def \JJ {{\rm E}}
\def \cA {{\cal A}} \def \rS   {{\rm S}} 
 \def \rA  {{\rm A}}
\def \l {\lambda}

\section{Introduction} 



Study of \sms  in \adst is of interest  for several reasons (see, e.g., \ci{Giombi:2017cqn,Beccaria:2019dws,Carmi:2018qzm}).  
Here we  will consider  correlators of elementary \sm fields  in Euclidean \adst 
with Poincare metric $ds^{2}=\frac{1}{\bz^{2}}(d\bt^{2}+d\bz^{2})$.
While in flat space  the    scattering  amplitudes of massless 
scalar fields in perturbative vacuum  are ambiguous  due to IR divergences  (see, e.g., \ci{Hoare:2018jim}) 
the  coordinate-space boundary correlators  in \adst  are  well-defined   and are constrained by 1d conformal invariance. 
One   interesting question  is  how the structure of these correlators  is further restricted  by  hidden symmetries 
  of the \sm  and   how to compute \adst  loop  corrections   in a way consistent with  these underlying symmetries. 

Since a  classical \sm  in  curved 2d space is Weyl-invariant (with the scalar field not transforming), defined on \adst  it is  formally the  same as 
on a half-plane $ds^{2}=d\bt^{2}+d\bz^{2}$, $\bz >0$. 
This  is    true  also at the quantum level if  the \sm  is  UV finite:  then  dependence on  the $\frac{1}{\bz^{2}}$  conformal factor  of the \adst metric 
 disappears. 
Compared  to a  generic  boundary CFT  set-up  here 
we are interested in (i) the standard \adst  (or Dirichlet)  boundary conditions 
$\vp(\bt, \bz)\big|_{  \bz\to 0  }  =  \bz^{\Delta}\,\sfPhi(\bt) + \cdots$  for an elementary  field with mass 
$m^{2}=\Delta(\Delta-1)$; 
(ii)  correlators  of  elementary  fields $\vp$  rather than  composite operators with good 2d conformal
 transformation properties. 
 The 1d  boundary   operators dual  to the massless   \sm  fields  with Dirichlet b.c. will thus have $\Delta=1$. 
 In contrast to the Liouville theory case 
 discussed in  \cite{DHoker:1983msr,Ouyang:2019xdd,Beccaria:2019stp} it  turns out 
 that    the  classical  
 2d conformal invariance of the bulk \sm theory does not  sufficiently   constrain the  structure of the  tree-level boundary correlators.
 For example, the tree-level boundary  four-point functions   are still non-trivial  log functions of 1d cross-ratio.
 An important
 difference  is that  while the \sm field is a scalar on which conformal symmetry acts trivially, 
  the Liouville field   transforms non-trivially   under the conformal transformations.\foot{\ 
 The theories  in flat space   and in \adst   correspond to different vacua \cite{DHoker:1983zwg,DHoker:1983msr,Zamolodchikov:2001ah}.
 The Liouville field in \adst has a  constant vacuum with the fluctuation field   with 
 $m^2=2$   and thus $\Delta=2$.   Its boundary correlators  are constrained by 1d Virasoro symmetry
 and thus are  exactly the  same as the  2d stress-tensor  correlators restricted to the  boundary
  \cite{Ouyang:2019xdd,Beccaria:2019stp}. This generalizes  also to the Toda theory   (see   also \ci{Beccaria:2019ibr,Beccaria:2019mev,Beccaria:2019dju}).
 }
 Similar  tree-level correlators  (containing logs)  were found also  for the fields of  the Nambu  action  in \adst  \ci{Giombi:2017cqn,Beccaria:2019dws}, 
 but they appear already in the  case of  2-derivative   \sm vertices.

 To study the role of additional \sm  symmetries   here we will  consider the example of the WZW  model
  \cite{Witten:1983ar,Novikov:1982ei}
 which has  an  infinite-dimensional Kac-Moody (KM) 
 symmetry  $g' = u(w)\, g\, v (\bar w)$, \ $w=\bt + i \bz$. 
  It appears  for the special   value of   the ratio 
 of the coefficients of the  principal chiral model (PCM)  and WZ terms in the action
  when the classical equations of motion admit a chiral decomposition (the resulting model  
 is then conformal  and KM invariant   also  at the quantum level). 
 Like the  Virasoro symmetry in the Liouville case   here the 
 KM symmetry 
 will impose rigid constraints on  the \adst boundary correlators of the elementary fields   $\vp_a $  parametrizing $g$. 
In particular, the KM symmetry   rules out the presence of log terms in the 
four-point correlators, both at the tree  and the quantum level.

As we  will  argue below, the \adst  boundary  correlators  of  the massless fields $\vp_a$   defined in the standard way as 
 \begin{align}
\la{1.1}
\langle\sfPhi_{a_1}(\bt_{1})  \cdots \sfPhi_{a_n}(\bt_{n})\rangle \equiv   \lim_{\bz_{i}\to 0}\prod_{i=1}^{n}\bz_{i}^{-\Delta}\,\langle \varphi_{a_1}(\bt_{1}, \bz_{1})\cdots \varphi_{a_n}(\bt_{n}, \bz_{n})
\rangle_{_{\text{AdS}_{2}}}\ , \qquad \Delta=1 \ , 
\end{align}
 are   constrained   by  the underlying   KM   symmetry  so that they are equal, up to a   universal  prefactor, 
   to the correlators 
 of the chiral  component of the  WZW current 
 $J^a (w) \sim  \tr ( t^a   \del_w  g g^{-1}), \ w=\bt + i \bz,$ restricted to the boundary. 
 This  
  is  formally equivalent 
 to the ``identification'' of the boundary operator associated to $\vp_a$   with the chiral component of the current $J_a(w\to \bt)$ 
 \be \la{1.2} 
  \sfPhi_{a}(\bt)\  \to\   \k\, J_a (w)\big|_{\bz \to 0}   \ , \ \ \qquad \qquad \k = \sqrt{ 2 \ov k} 
   \ .  \ee
Here $k$ is the  WZW level. 
  For comparison, in the Liouville theory case the role of the  $\Delta=1$   current $J\equiv J_w $  (the generator of KM symmetry) is played by the 
$\Delta=2$  chiral stress tensor  $T\equiv T_{ww}$  (the generator of the Virasoro  symmetry)\foot{\ In the Liouville (or Toda) case the 
Virasoro  symmetry becomes realized  as a reparametrizations of the boundary and thus completely  fixes   the 
structure of the  correlators modulo overall powers of the  coordinate-independent factor $\k$.}
  and  the proportionality coefficient was 
$\k  = - 4 \sqrt{ c-1 \ov 6 c^2}$   where $c=1+ 6 ( b^{-1} + b)^2$  is the  Liouville  central charge 
\ci{Beccaria:2019stp}.\foot{\ This 
close analogy may not be accidental given that  the Liouville theory may be obtained by a Hamiltonian reduction from the $SL(2)$ WZW  model \ci{ORaifeartaigh:1991dbm,Feher:1992yx}.}
 In the  WZW   case  the KM symmetry   implies the Virasoro symmetry  but is much  stronger: as already mentioned above, 
 the boundary correlators in  conformal \sms  that do not have an  extra  KM symmetry    have  much more   complicated 
 structure.~\foot{\ The key point is that  the elementary \sm field transforms non-trivially  under the  KM symmetry
 (like the Liouville field  was  transforming 
 under the conformal symmetry).  
 Note  also that the  simplification of the form of boundary correlators in the case  of  KM  symmetry 
  is analogous to what happens in the  AdS/CFT  examples when   the bulk  theory   has higher  symmetry  thus constraining also 
 the  correlators of the dual boundary CFT.  An  example  is provided   by the vectorial AdS/CFT   where 
 the symmetry in question is a higher spin symmetry \ci{Klebanov:2002ja,Maldacena:2011jn}.
  } 

Explicitly, the standard OPE relation for the  chiral   components of the  KM current (see, e.g.,  \ci{DiFrancesco:1997nk})
\be\la{1.3}
J^a(w) J^b(w') \sim \frac{k\delta^{ab}}{(w-w')^2} + \frac{  {  f}^{abc } J^c(w)}{w-w'} + \cdots~,
\ee
determines all higher current correlators to be given by 
\begin{align}\la{1.4}
\EV{J^{a_1}(w_1)\, J^{a_2} (w_2)} &=\frac{k\,\delta^{a_1 a_2}} {(w_1-w_2)^2} \ , \\
\EV{J^{a_1} (w_1)\, J^{a_2} (w_2)\, J^{a_3} (w_3)} &=\frac{k\, f^{a_1 a_2 a_3}}{w_{12}w_{13}w_{23}}\ , 
\la{1.5}\\ \la{1.6}
\EV{J^{a_1}(w_1)\, J^{a_2}(w_2)\, J^{a_3}(w_3)\, J^{a_4}(w_4) } 
&=\frac{k^2\, \delta^{a_1 a_2 } \delta^{a_3 a_4 } }{ w_{12}^2 w_{34}^2}
+\frac{k\, f^{a_1 a_2 b}f^{a_3 a_4 b }}{ w_{12}w_{34}w_{23}w_{24} }
+\PBK{2\leftrightarrow 3 }
+\PBK{ 2\leftrightarrow 4} \ . 
\end{align} 
Below we  will     explicitly reproduce \rf{1.4},\rf{1.5},\rf{1.6}  with $w_i\to \bt_i$
 as the expressions for the   boundary correlators  of the WZW fields 
\rf{1.1}  computed in the  $1/k$ perturbation theory  in \adst  
with the identification \rf{1.2}. 

A  semiclassical  argument  of why the  boundary correlators of $\vp_a$  are related to  the restriction of the current correlators to the  boundary of half-plane  can 
be given as follows (for a similar  though more involved argument in the Liouville theory case see  \ci{Beccaria:2019stp}). 
Starting with   the expression 
  $J^a \sim  \tr ( t^a  \del_w  g g^{-1} ) \to   \del_w \vp^a +  \OO(\vp^2) $ (where   $\del_w = \ha ( \del_\bz - i \del_\bt)$) 
and using the  boundary  condition
  $\vp_a({\bz}, \bt) \big|_{\bz\to 0} \to  {\bz}\,  \sfPhi_a (\bt) + ...$
we find that  (up to an overall normalization constant)  
 $J^a \big|_{\bz\to 0}  \to 
   \sfPhi^a $.

\

To  demonstrate  the correspondence \rf{1.2}   we shall  start   in section~\ref{sec2}   with 
the example of the 
$SL(2,\mathbb R)$ WZW model on \adst and compute   boundary correlators  of its fields in the  leading tree-level  approximation. 
We shall also  consider the corresponding PCM$_q$  theory
(i.e. the PCM with a WZ term with  coefficient  $\propto q$),  
and show that   the four-point correlators  simplify (with logs  of coordinates
  cancelling out) 
and thus can be matched with  the correlators  of the chiral currents only  at the WZW  point ($q^2=1$)  when  the model has an extra KM symmetry. 
In section~\ref{sec3} we shall  repeat the computation of the tree boundary correlators    for  a   generic \sm   including the case of    PCM$_q$  for an  arbitrary  group $G$. 

In section~\ref{sec4}  we shall   test   the relation between   the  boundary correlators of the  WZW  fields  and the chiral  currents \rf{1.2} 
beyond  the   classical (large $k$)  limit  by   computing the   one-loop corrections to the two-point and three-point  boundary correlators. 
Like in similar  computations  in the Liouville  and Toda theories in \adst 
 \cite{Beccaria:2019stp,Beccaria:2019mev,Beccaria:2019dju}   this requires an explicit evaluation of loop  integrals  in \adst
 which is subtle  in the present case  of the  \sm  theory with two derivatives in the  vertices.  We shall  argue  that 
 there exists a  particular computational scheme in which the  one-loop terms in the  WZW   field   boundary correlators   vanish, 
  implying that 
 the proportionality coefficient $\kappa$ in \rf{1.2} does not receive   $1/k$   correction  and thus  its expression in \rf{1.2} is expected to be exact. 
 
It is interesting to note that   while in flat space the   scattering  amplitudes   for the massless   WZW  fields 
vanish   \ci{Figueirido:1988ct,Hoare:2018jim} their   coordinate-space   boundary  correlators  in \adst  are non-vanishing. 
Their structure, however,  is simple being dictated by the  KM symmetry. 
One may wonder  if with some natural definition of the   S-matrix  in AdS  they  may   actually correspond to 
 trivial scattering in \adst or on half-plane. We  will address  this question in section~\ref{sec5}. There is a close analogy   with what 
  happens in the Liouville theory \ci{DHoker:1983msr}  where  the full  quantum S-matrix was argued to be trivial 
\ci{Thorn:1983qr,Yoneya:1984dd}. We shall discuss the idea  of defining \adst scattering amplitudes  by Fourier 
transform  of boundary correlators or using the prescription  of \ci{DHoker:1983msr} 
(cf. \ci{Giddings:1999qu})    and argue that this  leads to trivial 
three-point scattering amplitudes also  in the present WZW case. 

Some concluding remarks will be made in section~\ref{sec6}. Appendix~\ref{app:notation}  will list our notation  and conventions. In Appendix~\ref{app:symm} 
we shall  discuss  a constraint   imposed    by global symmetry  on boundary two-point functions in the   $SL(2,\mathbb R)$  WZW model.
In Appendix~\ref{appC} we shall revisit  the computation of the one-loop corrections to the two-point  boundary correlators 
in  $SL(2, \mathbb R)$    WZW model  using an alternative 
  form of the action  and   emphasizing   some subtle scheme-dependence issues.

\iffa 
The advantage of considering AdS space instead of flat space is that massless scattering amplitudes are problematic
in the latter, see \cite{Hoare:2018jim} or \cite{Donahue:2018bch}. Hopefully,  in AdS (equivalent to half space if Weyl
invariance holds) it could be possible to study the connection between integrability and factorization in massless
models like in massive case.
\fi

\section{
Boundary correlators in $SL(2,\mathbb R)$    WZW model   on \adst} \la{sec2}

To  demonstrate  the correspondence \rf{1.2}   we shall first consider the example of the 
$SL(2,\mathbb R)$ WZW model and compute  its boundary correlators on \adst  in the  leading-order (tree) approximation. 
It is useful to view  this WZW  model as a special   case  of the PCM$_q$,  i.e. the  principal chiral model with an additional WZ term.
This   allows  one  to investigate the consequences of the  Kac-Moody symmetry  appearing at the WZW point  
for the   structure of the  boundary correlators.

\subsection{Action} 


The action for the PCM$_q$   may be written as 
\be\label{2.1}
S= \frac{1}{2\pi\l^2}\,\Big[ -\tfrac{1}{2}\, \int_\Sigma d^2 x\, \textrm{Tr}(g^{-1}\partial_{\mu}g\; g^{-1}\partial^{\mu}g) 
+\tfrac{i}{3} q  \int_{\bf B^3} \textrm{Tr}(g^{-1}dg\wedge g^{-1}dg\wedge g^{-1}dg) 
\Big]~,\qquad    q= \ha k \l^2 \ , 
\ee
where $k$ is the  coefficient of the WZ term, 
 $\Sigma $ is a Riemann surface and  ${\bf B}^3$ is the 3d  extension of $\Sigma$ such that  
$\p{\bf B}^3=\Sigma$. When
\be \la{21}
\lambda = \sqrt{\tfrac{2}{|k|}}
 \ , \qquad {\rm i.e.}  \qquad   q=\sign k = \pm 1   \ , \ee
 the action (\ref{2.1}) reduces to the WZW model  action. 

Assuming $k > 0$,  a  generic $SL(2,\mathbb R)$ group element may be represented  in the Gauss  decomposition  form     (see,  e.g.,  \cite{Gerasimov:1990fi})
\be
\la{2.2}
g(x) = \begin{pmatrix} 1 & \psi \\ 0 & 1 \end{pmatrix}
\begin{pmatrix} e^{- \frac{\lambda}{\sqrt 2} \phi} & 0 \\ 0 & e^{ \frac{\lambda}{\sqrt 2}\phi} \end{pmatrix}
\begin{pmatrix} 1 & 0 \\ \frac{\lambda^2}{2}\tilde \psi  \quad & 1 \end{pmatrix}.
\ee

Then the action \rf{2.1}  (written on generic  curved 2-space with metric  $\sf g$)  becomes 
\be\label{2.3}
S=\frac{1}{4\pi} \int d^2 x \sqrt{\sf g} \Big[{\sf g}^{\mu\nu }\p_\mu \phi \p_\nu \phi   
+  e^{ b\phi}   \big( {\sf g}^{\mu\nu } +  i q\epsilon^{\mu\nu} \big)   \p_\mu \psi \p_\nu \tilde \psi      \Big] ~, \qquad \ \ 
b\equiv \sqrt 2  \lambda \ . 
\ee
Here $\epsilon^{\mu\nu} = {\varepsilon^{\mu\nu}\ov  \sqrt{\sf g}}$ is   the standard antisymmetric tensor. 
This  action may be interpreted as that of a 
 $\sigma$-model with  AdS$_{3}$  target space  and  particular 
$B$-field coupling.

 Specializing  to the 
$q=1$ WZW point  and  the Euclidean 
 \adst     background 
 (see Appendix~\ref{app:notation} for our notation and conventions)  
\be\la{2.4} 
ds^2 =  \frac{d\sft^2 +d\sfz^2 }{\sfz^2}=  -4 \frac{ dw d\bar w }{(w-\bar w)^2} , \qquad w=\sft+i\sfz,\quad\bar w=\sft-i\sfz,\qquad 
\sfz>0~,
\ee
we  
get the following expression   for the  corresponding $SL(2, \mathbb R )$ WZW  action 
\be
\la{2.5}
S=   \int \rmd^2 w \big(  \p\phi \bar\p\phi   +   e^{b\phi}   \p \psi \bar\p \tilde\psi   \big)  
=  \int \rmd^2 w \big(     \p\phi \bar\p\phi   +      \p \psi \bar\p \tilde\psi  +  b\,\phi  \p \psi \bar\p \tilde\psi
+   \tfrac{1}{4}   b^2  \phi^2  \p \psi \bar\p \tilde\psi   + \cdots  \big), \ 
 \   \ b= \tfrac{2}{\sqrt k}.
\ee
As the conformal factor of the metric decouples,   this  is  formally the same as the WZW action  on a    flat  half-plane $\sfz>0$. 
However, it will be useful to phrase  the  computation of the boundary correlators in the \adst language.

We shall assume that  the massless  fields 
 $\phi,\psi,\tilde\psi$  are subject to the standard (Dirichlet) boundary conditions   and thus they should be dual to 
 the boundary operators  with dimension $\Delta=1$, i.e. 
the asymptotic expansion of these fields near  $\sfz=0$  is 
\be
\la{2.6} \sfz\to 0\ : \ \ \ \ \ 
\phi(\sft,\sfz) =\sfz \, \sfPhi(\bt)+\cdots, \qquad \psi(\sft,\sfz) =\sfz\, \sfPsi(\bt)+\cdots, \qquad \tilde\psi(\sft,\sfz) =\sfz\, \widetilde{\sfPsi}(\bt)+\cdots~.
\ee
Our aim will   be   to  compute the tree level   boundary correlation functions  (\ref{1.1})  for the fields in \rf{2.5}   and then match them with 
the correlators of KM currents.

\subsection{Propagators}

The bulk-to-bulk  propagator of a massless scalar   in AdS$_2$ with   a  standard normalization  
$\ha \int d^2 w \del^\mu \phi \del_\mu \phi$  is 
\be\la{2.7} 
G_{\Delta=1}(\eta)=-\frac{1}{4\pi} \log \eta~,
\ee
where the geodesic distance is defined as
\be
\la{2.8}
 \eta =\frac{u}{u+2}, \qquad\qquad  u=\frac{(\sft-\sft')^2+(\sfz-\sfz')^2}{2\sfz \sfz' }~.
\ee
Hence, for the field $\phi$ in \rf{2.5}  we have (cf. \rf{A.7})\footnote{\ Note that   $w,w'$ in the propagators   are labels of   the points on half-plane:
 the  propagators may also  depend on the anti-holomorphic coordinates $\bar w, \bar w'$ 
 but we  do not indicate  this  to  simplify the  notation. 
}
\be\label{2.9}
g_{\phi\phi} (w,w')=\EV{\phi(w)\phi(w')}=
\begin{tikzpicture}[line width=1 pt, scale=0.8, baseline=-0.1cm,decoration={markings, mark=at position 0.53 with {\arrow{>}}}]
\coordinate (A1) at (-1,0);        \coordinate (A2) at (1,0); 
\node[above] at (A1) {$w$};   \node[above] at (A2) {$w'$};
\draw (A1)--(A2);
\draw[black,fill] (A1) circle(0.05);\draw[black,fill] (A2) circle(0.05);
\end{tikzpicture}
=2\pi G_{\Delta=1}(\eta)\equiv g(\eta)
= -\frac{1}{2} \log \eta(w,w')~.
\ee
The  bulk-to-bulk  propagator of the pair of fields  $\psi, \tilde \psi $ is similarly
 \be\la{2.10}
g_{\psi\tilde\psi} (w,w') =
\begin{tikzpicture}[line width=1 pt, scale=0.8, baseline=-0.1cm,decoration={markings, mark=at position 0.53 with {\arrow{>}}}]
\coordinate (A1) at (-1,0);        \coordinate (A2) at (1,0); 
\node[above] at (A1) {$w$};   \node[above] at (A2) {$w'$};
\draw[postaction={decorate}] (A1)--(A2);
\draw[black,fill] (A1) circle(0.05);\draw[black,fill] (A2) circle(0.05);
\end{tikzpicture}
=\EV{\psi(w) \tilde\psi(w')}
=2  g (w,w') =-  \log \eta(w,w')~. 
\ee
Given the structure of  the perturbative (small $b$ or large $k$) 
expansion in (\ref{2.5}),  it is useful also  to quote  the  propagators for the differentiated fields
\beqn
g_{\p\psi \bar \p \tilde\psi} (w,w') &=&\EV{\p\psi(w) \bar\p\tilde\psi(w')}=
\begin{tikzpicture}[line width=1 pt, scale=0.8, baseline=-0.1cm,decoration={markings, mark=at position 0.53 with {\arrow{>}}}]
\coordinate (A1) at (-1,0);        \coordinate (A2) at (1,0); 
\node[above] at (A1) {$\p\psi(w)$};   \node[above] at (A2) {$\bar\p \tilde{\psi}(w')$};
\draw[postaction={decorate}] (A1)--(A2);
\draw[black,fill] (A1) circle(0.05);\draw[black,fill] (A2) circle(0.05);
\end{tikzpicture} \nonumber
\\&=& \label{2.11}
\p_w \bar\p_{w'} g_{\psi\tilde\psi}(w,w')
=\p_w \Big( \frac{1}{\bar w-\bar w'} -\frac{1}{w-\bar w'} \Big) 
=\frac{1}{(w-\bar w')^2}+\pi\delta^{(2)} (w-w')~, \qquad
\eeqn
where we used the relations (\ref{A.6}).
The $\delta$-function piece   here will be important to account for  below. 
 
 To compute the boundary correlators, we will also need the bulk-to-boundary propagators
\begin{align}
\la{2.12}
\gb_{\phi\phi} (\sft ; w' ) &= \lim_{\sfz\to 0 } \frac{1}{\sfz} g_{\phi\phi} (\sft,\sfz ;\sft' ,\sfz' )
= \frac{2\sfz' }{ (\sft'-\sft)^2 + \sfz'{}^2}
 = \frac{-i  }{  \sft-w'  }+ \frac{ i  }{  \sft- \bar w'  }
\equiv \gb  (\sft ; w' )~,\notag  \\
\gb_{\psi\bar\p{\tilde\psi}} (\sft;w')&=\lim_{\sfz\rightarrow 0} \frac{1}{\sfz} \EV{\psi(w) \bar\p\tilde\psi (w')} 
 =\frac{2 i}{(\sft-\bar w' )^2}
 =\bar \p_{w'} \frac{2 i}{(\sft-\bar w' ) }
=2\; \bar\p ' \gb  (\sft ; w' )~, \notag \\
\gb_{\tilde\psi  \p{ \psi}} (\sft;w')&=\lim_{\sfz\rightarrow 0} \frac{1}{\sfz} \EV{\tilde\psi(w) \p\psi(w')} 
 =\frac{- 2i}{(\sft-  w' )^2}
=\p_{w'} \frac{-2 i}{(\sft-  w' ) }
=2 \; \p ' \gb  (\sft ; w' )~.
\end{align}

\subsection{Tree-level boundary correlation functions} 


\subsubsection{Two-  and three- point functions }

Considering the  boundary-to-boundary case of the  propagators \rf{2.9},  we get  the following two-point functions (using the notation in \rf{1.1}) 
\be\label{2.13}
\langle\sfPhi(\bt_{1})\sfPhi(\bt_{2})\rangle = \frac{2 }{\bt_{12}^{2}}~,\qquad\qquad
\langle\sfPsi(\bt_{1})\widetilde{\sfPsi} (\bt_{2})\rangle = \frac{4 }{\bt_{12}^{2}}~, 
\qquad \qquad \sft_{ij}\equiv \sft_i -\sft_j \ . 
\ee
These  have the same   form as the boundary restriction  of \rf{1.4}. 


The only non-zero  three-point function is $ \EV{\sfPhi \sfPsi \widetilde{\sfPsi} }$, which, at the tree level (leading order in $1/k$), is computed by 
the Witten diagram\footnote{\ Here the dashed circle represents the boundary of   AdS$_2$ and  solid lines 
 represent the propagators of  the corresponding  fields  in  the bulk. }
\be
 \begin{tikzpicture}[line width=1 pt, scale=0.6, rotate=0,baseline=-0.1cm,decoration={markings, mark=at position 0.53 with {\arrow{>}}}]
\coordinate (A1) at (90:2);  \coordinate (A2) at (210:2);  \coordinate (A3) at (-30:2);
\coordinate (B1) at (90:1);  \coordinate (B2) at (210:1);  \coordinate (B3) at (-30:1);
\draw[dashed] (0,0) circle (2);
\draw (A1)--(0,0); \draw[postaction={decorate}]  (A2)--(0,0); \draw[postaction={decorate}]   (0,0)--(A3);
\draw[fill=black] (A1) circle (0.1); \draw[fill=black] (A2) circle (0.1); \draw[fill=black] (A3) circle (0.1); 
\node[above] at (A1) {$\sfPhi(\sft_1) $};
\node[left] at (A2) {$\sfPsi(\sft_2) $};
\node[right] at (A3) {$\widetilde{\sfPsi}(\sft_3) $};
\end{tikzpicture} ~. \la{215}
\ee
We have
\be\la{2.15}
 A_3(\sft_1,\sft_2,\sft_3)\equiv 
 \EV{\sfPhi(\sft_1)  \sfPsi(\sft_2) \widetilde{\sfPsi}(\sft_3)  }=
 -b \int {\rmd}^2 w  \, \gb_{\phi\phi}(\sft_1, w)   \gb_{\psi\bar\p{\tilde\psi}} (\sft_2, w) \gb_{\tilde\psi  \p{ \psi}} (\sft_3, w ) ~.
\ee
Using the propagators in~(\ref{2.12}), we get
\be\la{217}
A_3(\sft_1,\sft_2,\sft_3)=-8b \int_0^\infty   d \sfz\  \int _ {-\infty}^\infty  d \sft \;  \frac{ \sfz}{\pi  \left[\left(\sft-\sft_1\right){}^2+\sfz^2\right]
 \left(-\sft+\sft_2+i \sfz\right){}^2 \left(\sft-\sft_3+i \sfz\right){}^2}~.
\ee
This integral can be done by  first computing the residues in the $\sft$ integration variable.  Integrating then  over    $\sfz$ one finds 
\be\label{2.17}
 \EV{\sfPhi(\sft_1)  \sfPsi(\sft_2) \widetilde{\sfPsi}(\sft_3)  }=
\frac{   4i\, b}{\sft_{12} \sft_{23}\sft_{31}}~\ . 
\ee
This   has again the same structure as the real-line limit of \rf{1.5}. 

\subsubsection{Four-point functions}

We now turn to the four-point functions the computation of which   is  little more involved. The only non-vanishing cases are the 
correlators $\EV{\sfPsi^2\widetilde{\sfPsi}^2}$ and  $\EV{\sfPhi^2 \sfPsi \widetilde{\sfPsi} }$. 

\paragraph{
$\EV{\sfPsi^2{\widetilde {\sf \Psi} ^2}}$: }

At tree level  this correlator  is given by the following Witten diagrams
\be\label{2.18}
 \begin{tikzpicture}[line width=1 pt, scale=0.8, rotate=0,baseline=-0.1cm,decoration={markings, mark=at position 0.53 with {\arrow{>}}}]
\coordinate (A1) at (135:2);  \coordinate (A2) at (45:2);  
\coordinate (A3) at (-45:2);   \coordinate (A4) at (-135:2);
\coordinate (B1) at (-1,0); \coordinate (B2) at (1,0);
\draw[densely dashed] (0,0) circle (2);
\draw[postaction={decorate}]   (A1)--(B1); \draw [postaction={decorate}]  (B1)--(A4); 
\draw[postaction={decorate}]   (A3)--(B2); \draw [postaction={decorate}]  (B2)--(A2); 
 \draw (B1)--(B2);
\draw[fill=black] (A1) circle (0.1); \draw[fill=black] (A2) circle (0.1); 
\draw[fill=black] (A3) circle (0.1); \draw[fill=black] (A4) circle (0.1); 
\node[left] at (A1) {$\sfPsi(\sft_1)$};
\node[right] at (A2) {$\widetilde{\sfPsi}(\sft_4)$};
\node[right] at (A3) {$ \sfPsi(\sft_3)$};
\node[left] at (A4) {$\widetilde{\sfPsi}(\sft_2)$};
\node[left] at (B1) {$w$};
\node[right] at (B2) {$w'$};
\end{tikzpicture} 
+
 \begin{tikzpicture}[line width=1 pt, scale=0.8, rotate=0,baseline=-0.1cm,decoration={markings, mark=at position 0.53 with {\arrow{>}}}]
\coordinate (A1) at (135:2);  \coordinate (A2) at (45:2);  
\coordinate (A3) at (-45:2);   \coordinate (A4) at (-135:2);
\coordinate (B1) at (  0,-1); \coordinate (B2) at ( 0,1);
\draw[densely dashed] (0,0) circle (2);
\draw[postaction={decorate}]   (A1)--(B2); \draw [postaction={decorate}]  (B1)--(A4); 
\draw[postaction={decorate}]   (A3)--(B1); \draw [postaction={decorate}]  (B2)--(A2); 
 \draw (B1)--(B2);
\draw[fill=black] (A1) circle (0.1); \draw[fill=black] (A2) circle (0.1); 
\draw[fill=black] (A3) circle (0.1); \draw[fill=black] (A4) circle (0.1); 
\node[left] at (A1) {$\sfPsi(\sft_1)$};
\node[right] at (A2) {$\widetilde{\sfPsi}(\sft_4)$};
\node[right] at (A3) {$ \sfPsi(\sft_3)$};
\node[left] at (A4) {$\widetilde{\sfPsi}(\sft_2)$};
\node[above] at (B2) {$w$};
\node[below] at (B1) {$w'$};
\end{tikzpicture}~.
 \ee  
We  can represent  the result as   
\be
\la{2.19}
\EV{\sfPsi(\sft_1)\widetilde{\sfPsi} (\sft_2)  \sfPsi(\sft_3)\widetilde{\sfPsi}(\sft_4) }
=A_4(\sft_1, \sft_2,\sft_3, \sft_4)+A_4(\sft_1, \sft_4,\sft_3, \sft_2)~,
\ee
where
\beqn
A_4(\sft_1, \sft_2,\sft_3, \sft_4)
&=&  (-b)^2\int {\rmd}^2 w{\rmd}^2 w' \,  \gb_{\psi\bar\p{\tilde\psi}} (\sft_1, w) \gb_{\tilde\psi  \p{ \psi}} (\sft_2; w )\;
g_{\phi\phi} (w,w')  \;
\gb_{\psi\bar\p{\tilde\psi}} (\sft_3, w' ) \gb_{\tilde\psi  \p{ \psi}} (\sft_4; w') \nonumber
\\&=&   2^4 b^2\;  \widetilde{\sf H}(\sft_2, \sft_1,\sft_4, \sft_3)~, \la{2.20}
\eeqn
\be\label{2.21}
\widetilde{\sf H}(\sft_1, \sft_2,\sft_3, \sft_4)
\equiv  \int {\rmd}^2 w{\rmd}^2 w' \;  \p \gb  (\sft_1, w) \bar\p \gb   (\sft_2; w )\;
g (w,w') \;
\p' \gb  (\sft_3, w') \bar\p' \gb   (\sft_4; w' ) ~.
\ee
To compute  this integral we  may first  integrate by parts at the vertex $w$, 
\begin{align}  \la{2.22}
\widetilde{\sf H}(\sft_1, \sft_2,\sft_3, \sft_4) &=
-  \int {\rmd}^2 w{\rmd}^2 w' \; \frac{- i}{\sft_1-w} \bar\p \gb   (\sft_2; w )\;
 \p  g (w,w') \;
\p' \gb  (\sft_3, w') \bar\p' \gb   (\sft_4; w' )  
\\&=
  \int d^2 w d^2 w' \,   \frac{ -  i   \sfz'}{\pi ^2 (-\bar w+\sft_2 )^2 (w-\sft_1 ) 
  (\sft_4-\bar w')^2 (-\sft_3+w')^2
   (w-w')    (w-\bar w') }~,\no 
   \qquad
\end{align}
where we ignored  2-derivative terms 
assuming
\be\label{2.23}
\p ' \bar \p'  \gb  (\sft ; w' ) =0.
\ee
Indeed,   possible terms with  $\delta^{(2)}(\sft-w')$   and its derivative   may be neglected here as they 
localize  the 
bulk point  to the boundary, and hence give zero  contributions  after  performing the bulk integral. 

The integral in (\ref{2.22}) can be evaluated by applying the residue theorem    
\be\label{2.24}
\widetilde{\sf H}(\sft_1, \sft_2,\sft_3, \sft_4) 
=\frac{  \log \big(  \frac{\sft_{12}\sft_{34}}{\sft_{14}\sft_{23}}\big)^2  + i\pi  \big(\sgn \sft_{12}+ \sgn \sft_{23}+ \sgn \sft_{34}+ \sgn \sft_{ 41} \big)    }{4\sft_{13}^2\sft_{24}^2}
-\frac{1}{2\sft_{12}\sft_{13}\sft_{24}\sft_{34}}~.
\ee
Then  from  (\ref{2.19})  we finally obtain  
\be\label{2.25}
\EV{\sfPsi(\sft_1)\widetilde{\sfPsi} (\sft_2)  \sfPsi(\sft_3)\widetilde{\sfPsi}(\sft_4)      }=
2^4 b^2 \; \widetilde{\sf H}(\sft_2, \sft_1,\sft_4, \sft_3) 
+  2^4 b^2 \; \widetilde{\sf H}(\sft_4, \sft_1,\sft_2, \sft_3)
=\frac{8b^2}{  \sft_{12}\sft_{23}\sft_{34}\sft_{41}}~.
\ee
Remarkably, all logarithmic  (and sign function) terms  present  in \eqref{2.24}  cancel out in the sum  of the two exchange Witten diagrams.
This 
cancellation is  crucial   in order to be  able to match \rf{2.25}  with  the  correlators of the KM currents  that are rational  functions of the 
differences of points (cf. \rf{1.6}).

 
\paragraph{
$\EV{\sfPhi^2 \sfPsi \widetilde{\sfPsi} }$:} 
 
 This correlator is   given by the sum of the following three diagrams 
 \be\label{2.26}
 \begin{tikzpicture}[line width=1 pt, scale=0.8, rotate=0,baseline=-0.1cm,decoration={markings, mark=at position 0.53 with {\arrow{>}}}]
\coordinate (A1) at (135:2);  \coordinate (A2) at (45:2);  
\coordinate (A3) at (-45:2);   \coordinate (A4) at (-135:2);
\coordinate (B1) at (  0, 1); \coordinate (B2) at (0,-1);
\draw[densely dashed] (0,0) circle (2);
\draw  (A1)--(B1); \draw   (B2)--(A4); 
\draw[postaction={decorate}]   (A3)--(B2); \draw [postaction={decorate}]  (B1)--(A2); 
 \draw[postaction={decorate}]  (B2)--(B1);
\draw[fill=black] (A1) circle (0.1); \draw[fill=black] (A2) circle (0.1); 
\draw[fill=black] (A3) circle (0.1); \draw[fill=black] (A4) circle (0.1); 
\node[left] at (A1) {$\sfPhi(\sft_1)$};
\node[right] at (A2) {$\widetilde\sfPsi(\sft_4)$};
\node[right] at (A3) {$ \sfPsi(\sft_3)$};
\node[left] at (A4) {$\sfPhi(\sft_2)$};
\node[above] at (B1) {$w$};
\node[below] at (B2) {$w'$};
\end{tikzpicture} 
+
 \begin{tikzpicture}[line width=1 pt, scale=0.8, rotate=0,baseline=-0.1cm,decoration={markings, mark=at position 0.53 with {\arrow{>}}}]
\coordinate (A1) at (135:2);  \coordinate (A2) at (45:2);  
\coordinate (A3) at (-45:2);   \coordinate (A4) at (-135:2);
\coordinate (B1) at (  0, 1); \coordinate (B2) at (0,-1);
\draw[densely dashed] (0,0) circle (2);
\draw (A1) to[out=-70, in=160] (B2); \draw   (A4) to[out=70, in=-160] (B1); 
\draw[postaction={decorate}]   (A3)--(B2); \draw [postaction={decorate}]  (B1)--(A2); 
 \draw[postaction={decorate}]  (B2)--(B1);
\draw[fill=black] (A1) circle (0.1); \draw[fill=black] (A2) circle (0.1); 
\draw[fill=black] (A3) circle (0.1); \draw[fill=black] (A4) circle (0.1); 
\node[left] at (A1) {$\sfPhi(\sft_1)$};
\node[right] at (A2) {$\widetilde{\sfPsi}(\sft_4)$};
\node[right] at (A3) {$ \sfPsi(\sft_3)$};
\node[left] at (A4) {$\sfPhi(\sft_2)$};
\node[above] at (B1) {$w$};
\node[below] at (B2) {$w'$};
\end{tikzpicture} 
+
 \begin{tikzpicture}[line width=1 pt, scale=0.8, rotate=0,baseline=-0.1cm,decoration={markings, mark=at position 0.53 with {\arrow{>}}}]
\coordinate (A1) at (135:2);  \coordinate (A2) at (45:2);  
\coordinate (A3) at (-45:2);   \coordinate (A4) at (-135:2);
\coordinate (B1) at (  0, 0); \coordinate (B2) at (0,-1);
\draw[densely dashed] (0,0) circle (2);
\draw  (A1)--(B1); \draw   (B1)--(A4); 
\draw[postaction={decorate}]   (A3)--(B1); \draw [postaction={decorate}]  (B1)--(A2); 
\draw[fill=black] (A1) circle (0.1); \draw[fill=black] (A2) circle (0.1); 
\draw[fill=black] (A3) circle (0.1); \draw[fill=black] (A4) circle (0.1); 
\node[left] at (A1) {$\sfPhi(\sft_1)$};
\node[right] at (A2) {$\widetilde{\sfPsi}(\sft_4)$};
\node[right] at (A3) {$ \sfPsi(\sft_3)$};
\node[left] at (A4) {$\sfPhi(\sft_2)$};
\node[above] at (B1) {$w$};
\end{tikzpicture} 
 \ee  
 It  can be written as 
 \be
 \la{2.27}
 \EV{\sfPhi(\sft_1)\sfPhi(\sft_2)  \sfPsi(\sft_3)\widetilde{\sfPsi}(\sft_4)      }=B_4(\sft_1, \sft_2,\sft_3, \sft_4)+B_4(\sft_2, \sft_1,\sft_3, \sft_4)
 +C_4(\sft_1, \sft_2,\sft_3, \sft_4) ~,
 \ee
  where the explicit form of  $B_4$ and $C_4$ is 
   \begin{align}\label{2.28}
B_4(\sft_1, \sft_2,\sft_3, \sft_4)  
&=b^2\int {\rmd}^2 w{\rmd}^2 w' \,  \gb_{\phi\phi} (\sft_1, w) \gb_{\phi\phi} (\sft_2; w' )\; 
g_{\p\psi \bar \p \tilde\psi} (w',w ) 
\;  \gb_{\psi\bar\p{\tilde\psi}} (\sft_3, w' ) \gb_{\tilde\psi  \p{ \psi}} (\sft_4; w ) \qquad
\qquad\notag \\&=
B^\text{reg}_4(\sft_1, \sft_2,\sft_3, \sft_4)  -C_4(\sft_1, \sft_2,\sft_3, \sft_4)  ~,
 \end{align}
    \be\la{2.29}
C_4(\sft_1, \sft_2,\sft_3, \sft_4)  
=-b^2\int {\rmd}^2 w \,  \gb_{\phi\phi} (\sft_1, w) \gb_{\phi\phi} (\sft_2; w ) 
\; \gb_{\psi\bar\p{\tilde\psi}} (\sft_3, w ) \gb_{\tilde\psi  \p{ \psi}} (\sft_4; w ) ~.
 \ee
 Here  $B_4^\text{reg}$ and $C_4$ are the contributions from the regular   
 and singular parts of the internal propagators in \eqref{2.11}, respectively  (note that   in (\ref{2.28})
 the singular $\delta$-function part in the propagator \rf{2.11} turns the exchange diagram into a 
 contact diagram).  Using the explicit form  of the propagators we get 
  \beqn
B^\text{reg}_4(\sft_1, \sft_2,\sft_3, \sft_4)&=&      \frac{16 b^2 }{\pi^2} \int d^2 w d^2 w' \,  
 \frac{ \sfz \sfz'     }
 {(\sft_4-\sft-i \sfz  )^2 (\sft_3-\sft'+i \sfz'  )^2   \Big[(\sft-\sft_1)^2+\sfz^2\Big] \Big[(\sft'-\sft_2)^2+\sfz'{}^2\Big]     }
\nonumber\\&&\qquad\qquad\qquad\quad
\times  \Big[-\sft+\sft'+i(\sfz+\sfz')\Big]^{-2  }\ , 
  \\
 C_4(\sft_1, \sft_2,\sft_3, \sft_4) &=&  \frac{-16b^2}{\pi}\int d^2 w \, \frac{\sfz^2}{(\sft_3-\sft+i\sfz)^2  (\sft_4-\sft-i\sfz)^2   
 \Big[ (\sft_1-\sft)^2+\sfz^2 \Big]  \Big[(\sft_2-\sft)^2+\sfz^2 \Big]   }~.
 \eeqn
Using  the same method as for the previous four-point function, we obtain 
 \beqn \label{2.32}
B^\text{reg}_4(\sft_1, \sft_2,\sft_3, \sft_4)&=&- \frac{2ib^2}{\sft_{23}^2 \sft_{14}^2} \Big[   i \log \Big(  \frac{\sft_{12}\sft_{34}}{\sft_{13}\sft_{24}} \Big)^2 
 +\pi \big(- \sgn \sft_{12}  +\sgn \sft_{13} -\sgn \sft_{24} + \sgn\sft_{34} \big)
  \Big] ~,
\\  
 C_4 (\sft_1, \sft_2,\sft_3, \sft_4) &=&2 b^2  \Big[-i \frac{   i \log   \big(  \frac{     \sft_{12}          \sft_{34}    }{     \sft_{14}          \sft_{23}    }  \big)^2
   +\pi  \text{ sgn} ( \sft_{12} )-\pi  \text{ sgn} ( \sft_{14} )+\pi  \text{ sgn} ( \sft_{23} )+\pi  \text{ sgn} ( \sft_{34} ) }{ \sft_{13}^2  \sft_{24}^2}
  \nonumber\\&&\qquad
 + i\frac{  -  i \log    \big(   \frac{    \sft_{12}          \sft_{34}    }{     \sft_{13}          \sft_{24}    }   \big)^2  +\pi   \text{ sgn} ( \sft_{12} )-\pi   \text{ sgn} ( \sft_{13} )+\pi   \text{ sgn} ( \sft_{24} )-\pi   \text{ sgn} ( \sft_{34} ) }{ \sft_{14}^2  \sft_{23}^2}
 \nonumber\\&&\qquad
 +\frac{2}{ \sft_{13}  \sft_{14}  \sft_{23}  \sft_{24}} \Big]~. \la{2.33}
   \eeqn
  Inserting these results into  (\ref{2.27}) gives\footnote{\ 
  It is easy to verify that  $B_4$ is related to $A_4$ in \rf{2.19} as   $B_4(\sft_1, \sft_2,\sft_3, \sft_4)=A_4  (\sft_1, \sft_4,\sft_3, \sft_2)$.
  This relation can be easily understood   using   integration by parts.
   Indeed, integrating by parts the cubic vertex, one can transfer the derivatives acting on the internal leg  to the external legs;
    then the first diagram in \eqref{2.26} reduces to the second diagram in~\eqref{2.18}.
  }
    \beqn\label{2.34}
 \EV{\sfPhi(\sft_1)\sfPhi(\sft_2)  \sfPsi(\sft_3)\widetilde{\sfPsi}(\sft_4)      }
 =B^\text{reg}_4(\sft_1, \sft_2,\sft_3, \sft_4)+B^\text{reg}_4(\sft_2, \sft_1,\sft_3, \sft_4)
 -C_4(\sft_1, \sft_2,\sft_3, \sft_4) 
 =-\frac{4 b^2}{\sft_{13} \sft_{14} \sft_{23} \sft_{24}}  \qquad
 \eeqn
 As in  the  case of the  correlator    $\EV{\sfPsi(\sft_1)\widetilde{\sfPsi} (\sft_2)  \sfPsi(\sft_3)\widetilde{\sfPsi}(\sft_4)      }$     in \rf{2.25}, both the logarithms and the sign functions  again  cancel out.

\subsection{Matching \adst boundary correlators with  correlators of  chiral currents }


  Let us now compare   the above boundary correlators with the   correlators of the chiral WZW  currents on the plane   restricted to the real line. 
The correlation functions of the currents are a direct consequence of the  KM  algebra \rf{1.3}. 
Adapted to  the $SL(2,\mathbb R)$    case  the  OPEs  of the   three currents  $(H, J^+, J^-)$  read (see, e.g., ~\cite{Gerasimov:1990fi})
\begin{align}
\la{2.35}
H(w) H(0) &\sim \frac{k}{w^{2}}~, \qquad
H(w) J^{\pm}(0) \sim \mp \,\frac{i\sqrt 2\,  J^{\pm}(0)}{w}~, \qquad J^{+}(w) J^{-}(0) \sim \frac{2k}{w^{2}}- \,\frac{2i\sqrt 2\,  H(0)}{w}~.
\end{align}
From (\ref{2.35}) we   conclude that:   (i)   the two-point functions  are 
\be\label{2.36}
\langle H(w_1) H(w_2)\rangle = \frac{k}{w_{12}^{2}}~,\qquad \langle J^{+}(w_1) J^{-}(w_2) \rangle = 
 \frac{2\,k}{w_{12}^{2}}~ ,  
\ee
(ii) the only non-vanishing three-point function is 
\be\label{2.37} 
\langle H(w_{1})\,J^{+}(w_{2})\,J^{-}(w_{3})\rangle = - 2\,\sqrt{2}\,\frac{i\,k}{w_{12}\,w_{13}\,w_{23}}~,
\ee
and (iii) the  non-trivial   four-point functions are 
\begin{align}\label{2.38}
\langle J^{+}(w_{1})\ J^{-}(w_{2})\ J^{+}(w_{3})\ J^{-}(w_{4})\rangle &= 
4\,k^{2}\,\Big(\frac{1}{w_{23}^{2}w_{14}^{2}}+\frac{1}{w_{12}^{2}w_{34}^{2}}\Big)
-\,\frac{8\,k}{w_{12}\,w_{23}\,w_{14}\,w_{34}}~,\notag \\
\langle J^{+}(w_{1})\ J^{-}(w_{2})\ H(w_{3})\ H(w_{4})\rangle &= 
2\,k^{2}\,\frac{1}{w_{12}^{2}\,w_{34}^{2}}
- \frac{4\,k}{w_{13}\,w_{14}\,w_{23}\,w_{24}}~.
\end{align}
These four-point functions are non-trivial in the sense that  in addition to the $k^{2}$ contribution 
they also   have  a term linear in $k$  (cf. \rf{1.6}).\footnote{\ 
For instance, a four-point function which is non-vanishing but trivial in the above 
sense  is
\be\notag 
\langle H(w_{1})\ H(w_{2})\ H(w_{3})\ H(w_{4})\rangle = 
\frac{k^{2}}{4} \Big( \,\frac{1}{w_{12}^{2}\,w_{34}^{2}}+
\frac{1}{w_{13}^{2}\,w_{24}^{2}}+
\frac{1}{w_{14}^{2}\,w_{23}^{2}}\Big)\ . 
\ee
}
Comparing to boundary correlators discussed above, 
the $k^2$ term  is  a counterpart  with disconnected \adst Witten diagram contribution, 
while the order $k$ term  corresponds to  connected  exchange  
and contact 
  contributions   due to non-trivial bulk interactions. 
In fact, it is   possible to establish  the precise matching  between  the  tree-level  
2-point and 3-point  boundary correlators   in 
\eqref{2.13},\eqref{2.17} and the  current correlators  \eqref{2.36},\eqref{2.37} restricted to the real  line  using the following   identification (cf. \rf{1.2}) 
 \be
\sfPhi = \kappa\,H,\qquad \sfPsi = \kappa\,J^{+}, \qquad \widetilde{\sfPsi}  = \kappa\,J^{-} \ , \qquad \qquad  
\kappa= \sqrt { 2 \ov k} = {b\ov \sqrt 2} 
  \ . \la{2.39}
\ee
A non-trivial consistency check is that  the   connected  4-point  boundary correlators \rf{2.25},\rf{2.35}  then also match with the   non-trivial 
order $k$ parts of the  4-current correlators in \rf{2.38}. 
The  fact that   there is  just  a single universal   proportionality  coefficient $\kappa$   follows  from the  global  group symmetry of the WZW model 
(this is true not only at the tree level  but also  to all orders in $1/k$).

\iffa 
Comparing the two and three-point   functions in \eqref{2.13}, \eqref{2.17} and \eqref{2.36}, \eqref{2.37}
gives 
\be 
\kappa_{\phi}^{2}\frac{k}{2} = 2, \qquad 
\kappa_{\psi}^{2} k= 4, \qquad
\kappa_{\psi}^{2}\kappa_\phi  (- i k )=-4ib =-4i\frac{2}{\sqrt k}  
\ee
that is
\be\label{2.40} 
\kappa_{\phi} =  \frac{2}{\sqrt k}~,  \qquad \kappa_{\psi}^{2} = \frac{4}{k}~.
\ee
It is easy to check that this also ensures the agreement of connected part of four-point   functions in \eqref{2.25}, \eqref{2.33} and  \eqref{2.38}. The relation $\kappa_{\psi}^{2} =\kappa_{\phi} ^2 $ 
can be shown to hold at quantum level and follows from the  symmetry of WZW model.
\fi 

   
\subsection{Boundary correlators in PCM$_q$ on AdS$_2$} 

Let us now go back to the  $SL(2,\mathbb R)$  principal chiral model with  a general coefficient $q$  of the WZ  term  in \rf{2.3} 
to emphasize   that its boundary correlators   have a     complicated   structure   already at the tree level (containing, in particular,   logarithmic terms
found also   in a similar \sm context in  \ci{Giombi:2017cqn,Beccaria:2019dws}). 
In  contrast to the Liouville   and Toda  theories   discussed in  \cite{Ouyang:2019xdd, Beccaria:2019stp, Beccaria:2019ibr}
  here  the  classical   conformal symmetry  of PCM$_q$ \rf{2.3}    is not enough 
to  sufficiently constrain the  boundary correlators.\foot{\ 
As we remarked in the Introduction, 
this  may be  related  to the  fact that  the \sm fields transform 
 as scalars (i.e. trivially)   under the conformal group. }
The correlators        simplify  precisely  at the WZW point $q^2=1$    and this may be   attributed to  the emerging   KM symmetry 
(that implies chiral decomposition in   flat space). 


The action   \eqref{2.3} in \adst   expanded in   powers of $b $ reads (cf. \rf{2.5})\foot{\ While the \adst  conformal factor decouples at the tree level, this 
will no longer be so  at the quantum level as this model will have  UV divergences  and thus conformal anomaly (assuming
 reparametrization-covariant regularization) 
 unless  $q^2=1$.}
\beqn
S&=&   \int \rmd^2 w \Big(  \p\phi \bar\p\phi   +  \tfrac{1+q}{2} e^{b\phi}   \p \psi \bar\p \tilde\psi +  \tfrac{1-q}{2} e^{b\phi}  \bar \p \psi\p \tilde\psi   \Big)  
\nonumber \\&=&
  \int \rmd^2 w \Big(     \p\phi \bar\p\phi   +      \p \psi \bar\p \tilde\psi  
  +  \tfrac{b(1+q)}{2}\,\phi  \p \psi \bar\p \tilde\psi+   \tfrac{b(1-q)}{2} \, \phi \bar  \p \psi  \p \tilde\psi
\nonumber \\& &\qquad\qquad
 +   \tfrac{b^2(1+q)}{8} {  \phi^2}  \p \psi \bar\p \tilde\psi    +  \tfrac{b^2(1-q)}{8}  {  \phi^2}    \bar\p \psi \p \tilde\psi   + \cdots  \Big)~.\la{2.40}
\qquad
\eeqn
Repeating the 
 calculation of the tree-level  four-point  $\EV{\sfPsi^2 \widetilde{\sfPsi}^2}$  correlator  we find (cf.  \rf{2.19})
 \be\label{2.41}
\EV{\sfPsi(\sft_1)\widetilde{\sfPsi} (\sft_2)  \sfPsi(\sft_3)\widetilde{\sfPsi}(\sft_4)      }= \mathcal  A_4(\sft_1, \sft_2,\sft_3, \sft_4)+\mathcal  A_4(\sft_1, \sft_4,\sft_3, \sft_2)~,
\ee
\begin{align}
\mathcal A_4(\sft_1, \sft_2,\sft_3, \sft_4)
&=
\big( \tfrac{1+q}{2}  \big)^2 A_4(\sft_1, \sft_2,\sft_3, \sft_4)+\big( \tfrac{1-q}{2}  \big)^2 A_4(\sft_2, \sft_1,\sft_4, \sft_3)
\notag \\ &
+\tfrac{1-q^{2}}{4}\,  \Big[A_4(\sft_2, \sft_1,\sft_3, \sft_4)
+A_4(\sft_1, \sft_2,\sft_4, \sft_3)\Big]~,\la{2.42}
\end{align}
where $A_4$ is given by \rf{2.20},\rf{2.21},\rf{2.24}. 
Similarly,  for the $\EV{\sfPhi^2 \sfPsi\widetilde{\sfPsi} }$  correlator   we get 
 \be\label{2.43}
 \EV{\sfPhi(\sft_1)\sfPhi(\sft_2)  \sfPsi(\sft_3)\widetilde{\sfPsi}(\sft_4)      }=\mathcal B_4(\sft_1, \sft_2,\sft_3, \sft_4)+\mathcal  B_4(\sft_2, \sft_1,\sft_3, \sft_4)
+\mathcal  C_4(\sft_1, \sft_2,\sft_3, \sft_4) ~,
 \ee
 \beqn
 \mathcal B_4(\sft_1, \sft_2,\sft_3, \sft_4)&=&\big(\tfrac{1+q }{2}  \big)^2 B_4(\sft_1, \sft_2,\sft_3, \sft_4)
 +\big(\tfrac{1-q }{2}  \big)^2 B_4(\sft_1, \sft_2,\sft_4, \sft_3)~,\la{2.44} 
\\ 
 \mathcal C_4(\sft_1, \sft_2,\sft_3, \sft_4) &=& \tfrac{1+q }{2}  C_4(\sft_1, \sft_2,\sft_3, \sft_4)
 + \tfrac{1-q }{2}   C_4(\sft_1, \sft_2,\sft_4, \sft_3)~,\la{2.45}
 \eeqn
 where $B_4$ and $C_4$  are given by \rf{2.28},\rf{2.29},\rf{2.32},\rf{2.33}. 
 We conclude that  these four-point functions  contain logarithmic  terms. These  cancel only at the WZW point $q^2=1$
 allowing one to relate  these boundary correlators to the ``connected''   part of the  correlators of the chiral WZW currents as  explained above.



\section{Tree-level boundary correlators in  generic   $\s$-model on \adst }  \la{sec3}

Let   us now consider the \adst boundary   correlators  in a general \sm   expanded near a constant background.
This includes, in particular, the case of a WZW  model for a general group $G$. 
We shall again  demonstrate  the cancellation of the logarithmic terms in the four-point correlators  at the WZW point 
and match them with the connected part of the  correlators of the chiral currents \rf{1.6}  restricted  the real line.


\subsection{Action}

Let us start with a general bosonic \sm  with coupling  functions $(G_{ab}, B_{ab})$ 
and  expand it  in normal coordinates near  the  origin using  
$
 G_{ab}(X)=\delta_{ab} -\frac13 R_{acbd}(0) X^c X^d+\mathcal O(X^3) 
 $
 and
 $
 B_{ab}(X)= B_{ab}(0)+  \frac13 H_{abc}(0) X^c+\mathcal O(X^2)
$.  Then  its Euclidean action  may be written as\foot{\ We ignore   the overall   coupling factor  or $1\ov \alpha'$ that can be absorbed into a  rescaling of $X^a$
and then appears in $R$ and $H$.} 
 \beqn\no
 S&=&\frac{1}{4\pi }\int d^2 x  \sqrt{\sf g}
\;
\Big[{\sf g}^{\mu\nu} G_{ab}(X) +i \epsilon^{\mu\nu} B_{ab}(X)\Big] \p_\mu X^a \p_\nu X^b
\\&=&\la{3.1}
\frac{1}{4\pi}\int d^2 x  \sqrt{\sf g}\;
\Big[
{\sf g}^{\mu\nu} \big(\delta_{ab} -\tfrac13 R_{acbd}(0) X^c X^d+\cdots\big) + \tfrac {i}{3} \epsilon^{\mu\nu}\big(H_{abc}(0) X^c +\cdots\big)
 \Big] \p_\mu X^a \p_\nu X^b~.
 \eeqn
 In what   follows we will  consider the leading terms in this action  parametrized as
 \be\label{3.2}
 S=\frac{1}{4\pi} \int d^2 x \sqrt{\sf g}  
\Big(\p_\mu X^a \p^\mu X^a -P_{abcd} X^a X^c \p_\mu X^b \p^\mu X^d
+i Q_{abc} \epsilon^{\mu\nu} X^a \p_\mu X^b \p_\nu X^c+\cdots\Big)~,
\ee
 where the constant real  coupling functions $P$ and $Q$ are given by 
  \be\la{3.3} 
 P_{abcd} \equiv  \tfrac13 R_{abcd}(0) ~, \qquad\qquad  Q_{ abc }\equiv  \tfrac{1}{3} H_{ abc }(0) ~.
 \ee
 Thus $Q$ is totally antisymmetric and $P$  has algebraic symmetries of the curvature 
\begin{align} \label{3.4}
&P_{abcd}=-P_{bacd}~, \qquad P_{abcd}=-P_{abdc}~, \qquad P_{abcd}=P_{cdab}~\ , \\
&\label{3.5}
P_{abcd}+P_{acdb}+P_{adbc}=0~.
\end{align}
To  account for the manifest  symmetry of the 4-vertex in \rf{3.2}  in $(a,c)$  it is useful to 
introduce also  the corresponding symmetrization of $P _{abcd}$
\begin{align}
&\label{3.6}
\tilde P _{abcd}\equiv \tfrac12 (P_{abcd}+ P_{c bad})=-\tfrac12 (P_{ bacd} +P_{ bcad})
= \tfrac12 (P_{ badc} +  P_{ dabc})~, \\
&\label{3.7}
\tilde P_{abcd}= \tilde  P_{ badc}~,\qquad 
\tilde  P_{abcd}= \tilde  P_{cbad}~,\qquad
\tilde  P_{abcd}= \tilde  P_{adcb}~.\qquad
\end{align}
Then  specifying to  the \adst   background   the action \rf{3.2} may be written  as (cf. \rf{2.5}) 
\be\label{3.8}
  S= \int \rmd^2 w 
  \Big(\p X^a \bar \p X^a - \tilde P_{abcd} X^a X^c \p  X^b \bar\p X^d
+Q_{abc}  X^a \p  X^b \bar \p  X^c+\cdots\Big)~.
\ee
The  action \rf{3.2}    represents  as a particular case 
  the expansion of the PCM$_q$ \rf{2.1}   for an arbitrary 
group $G$. 
Let us normalize the   generators $\{ t_a\}$ and the invariant bilinear form of the Lie algebra  of  $G$  as\foot{\ We assume that     $t_a$ are Hermitian and thus the  structure constants  $f_{ab}{}^c$ are purely imaginary. 
The group indices are raised or lowered by $\delta_{ab}$, implying that   $f_{abc} =f_{ab}{}^c  $ 
is fully anti-symmetric.  Repeated group indices are summed over, regardless of their positions.}  
\be\label{3.9}
[t_a,t_b]=f_{ab}{}^c t_c~, \qquad\qquad  \tr (t_a t_b)=  \delta_{ab} ~.
\ee
 Then choosing the parametrization  of the group field  as\footnote{\ Notice that  this  parametrization is different from (\ref{2.2}) used in the $SL(2, \mathbb R)$ case.
}
  $g=e^{-i \lambda t_a X^a}$   we find that in the \pcmq   case 
\be\label{3.10}
\text{PCM}_q: \quad 
P_{abcd} ={\sf p} f_{abe} f_{cd  e} ~, \qquad Q_{abc}={\sf q} f_{abc}~, \qquad \ \ \ {\sf p}=-\tfrac{1}{12} \lambda^2, \ \  \ \ \ 
{\sf q}= - \tfrac13  i \lambda q~\ . 
\ee
The WZW   theory corresponds to the   choice \rf{21}, i.e.   $ {\sf q}=- \sign(k)   \tfrac13  i  \l, \ \  \l=  \sqrt{\frac{2}{|k|}}$.
In what follows we shall assume  that $k >0$. 



\subsection{Tree-level   \adst   boundary  correlation functions}

The fields $X^a$  in \eqref{3.8} are massless  and thus,  assuming the  Dirichlet boundary conditions, 
we have  (cf. (\ref{2.6}))
\be\la{3.11}
   X^a(\sft, \sfz )\big|_{\sfz\to 0} =\sfz\; \sfX^a(\sft)+\cdots~\ . 
\ee 
  They should  correspond to   the boundary operators with dimension $\Delta=1$. 
As in (\ref{2.9}), their  bulk-to-bulk  \adst propagator  is given by 
\be\la{3.12}
g_{ab} (w,w')=\EV{X_a(w)X_b(w')}
=\delta_{ab} \; g(\eta)
= -\frac{1}{2}\delta_{ab} \log \eta(w,w')\ ,
\ee
while the bulk-to-boundary  propagator is 
\be\la{3.13}
\gb_{ab}   (\sft ; w' )=\lim_{\sfz\to 0 } \frac{1}{\sfz} g_{ab} (\sft,\sfz ;\sft' ,\sfz' )
=\delta_{ab} \frac{2\sfz' }{ (\sft'-\sft)^2 + \sfz'{}^2} 
=\delta_{ab} \; \gb(\sft, w') ~.
 \ee
  Then the boundary  two-point function is (cf. \rf{2.13}) 
  \be\label{3.14}
 \EV{\sfX_a(\sft_1) \sfX_b(\sft_2)} =\frac{2\delta_{ab}}{ \sft_{12}^2}~.
 \ee
Starting with \rf{3.8}   it is  straightforward  also to compute the three-point function 
\be\label{3.15}
 \EV{{\sfX}_a(\sft_1)\sfX_b(\sft_2)\sfX_c(\sft_3)  }= 
 \begin{tikzpicture}[line width=1 pt, scale=0.6, rotate=0,baseline=-0.1cm,decoration={markings, mark=at position 0.53 with {\arrow{>}}}]
\coordinate (A1) at (90:2);  \coordinate (A2) at (210:2);  \coordinate (A3) at (-30:2);
\coordinate (B1) at (90:1);  \coordinate (B2) at (210:1);  \coordinate (B3) at (-30:1);
\draw[dashed] (0,0) circle (2);
\draw (A1)--(0,0); \draw (A2)--(0,0); \draw   (0,0)--(A3);
\draw[fill=black] (A1) circle (0.1); \draw[fill=black] (A2) circle (0.1); \draw[fill=black] (A3) circle (0.1); 
\node[above] at (A1) {$\sfX_a(\sft_1) $};
\node[left] at (A2) {$\sfX_b(\sft_2) $};
\node[right] at (A3) {$ \sfX_c(\sft_3) $};
\end{tikzpicture}  
= \frac{  6i Q_{abc}}{\sft_{12} \sft_{ 23} \sft_{13 } }~.
\ee
The  connected tree-level 
 four-point function  receives
contributions from both  exchange diagrams and  contact diagrams
\be\label{3.16}
 \EV{{\sfX}_a(\sft_1)\sfX_b(\sft_2)\sfX_c(\sft_3)\sfX_d(\sft_4)}= 
\sG^\text{exch}_{abcd} (\sft_1,\sft_2,\sft_3,\sft_4)+\sG^\text{cont}_{abcd} (\sft_1,\sft_2,\sft_3,\sft_4)~.
\ee

\paragraph{Exchange diagrams.} The exchange  part contains  contributions of the three  different channels
\be\label{3.17}
\sG^\text{exch}_{abcd}  (\sft_1,\sft_2,\sft_3,\sft_4)
=G_{abcd}^\text{exch}(\sft_1,\sft_2,\sft_3,\sft_4)+G_{acbd}^\text{exch}(\sft_1,\sft_3,\sft_2,\sft_4)
+G_{adcb}^\text{exch}(\sft_1,\sft_4,\sft_3,\sft_2)~. \qquad
\ee
As these are related by permutations (crossing), we 
 only need to compute one of them
\beqn\la{3.18}
G_{abcd}^\text{exch}(\sft_1,\sft_2,\sft_3,\sft_4)=
   Q_{abe}Q_{cde}
 \begin{tikzpicture}[line width=1 pt, scale=0.6, rotate=0,baseline=-0.1cm,decoration={markings, mark=at position 0.53 with {\arrow{>}}}]
\coordinate (A1) at (135:2);  \coordinate (A2) at (45:2);  
\coordinate (A3) at (-45:2);   \coordinate (A4) at (-135:2);
\coordinate (B1) at (  0, 1); \coordinate (B2) at (0,-1);
\draw[densely dashed] (0,0) circle (2);
\draw  (A1)--(B1); \draw   (B2)--(A4); 
\draw   (A3)--(B2); \draw  (B1)--(A2); 
 \draw  (B2)--(B1);
\draw[fill=black] (A1) circle (0.1); \draw[fill=black] (A2) circle (0.1); 
\draw[fill=black] (A3) circle (0.1); \draw[fill=black] (A4) circle (0.1); 
\node[left] at (A1) {$\sfX_a(\sft_1)$};
\node[right] at (A2) {$\sfX_b(\sft_2)$};
\node[right] at (A3) {$ \sfX_c(\sft_3)$};
\node[left] at (A4) {$\sfX_d(\sft_4)$};
\node[above] at (B1) {$w$};
\node[below] at (B2) {$w'$};
\end{tikzpicture} ~.
\eeqn
 Note that using integration by parts, one can always  arrange so that 
  the derivatives in  the cubic vertex in \eqref{3.8}   act only on the two external legs.\footnote{\ Note that 
 $X^a  \big( \p X^b \bar\p X^c -\p X^c \bar\p X^b  \big)$ is fully anti-symmetric in $a,b,c$  up to a total derivative. }
 Then  the 6 terms in the cubic vertex can be written as 
 \be
 \la{3.19}
 K(\sft_1,\sft_2,w)=3 \p \gb(\sft_1, w) \bar\p  \gb(\sft_2, w) -3 \bar\p \gb(\sft_1, w) \p  \gb(\sft_2, w) ~,
 \ee
 where the two terms arise from the  two ways of acting  by derivative on the external legs and the factor  of 
 3 comes  from  rearranging  other similar cubic terms. Using (\ref{3.19}), we  find for  the exchange diagram  
  \beqn\label{3.20}
 G_{abcd}^{\text{exch}}(\sft_1,\sft_2,\sft_3,\sft_4) 
&=& Q_{abe } Q_{cde } \int \rmd^2 w\; \rmd^2 w'\;     K(\sft_1,\sft_2,w) \ g(w,w') \  K(\sft_3,\sft_4,w)
\nonumber \\&=&
9 Q_{ abe } Q_{ cde }  {\sf H} (\sft_1,\sft_2,\sft_3,\sft_4)~,
\eeqn
\be
\la{3.21}
 {\sf H} (\sft_1,\sft_2,\sft_3,\sft_4) =  \tilde{\sf H}(\sft_1,\sft_2,\sft_3, \sft_4) -  \tilde{\sf H}(\sft_1,\sft_2,\sft_4, \sft_3)-  \tilde{\sf H}(\sft_2,\sft_1,\sft_3, \sft_4)+  \tilde{\sf H}(\sft_2,\sft_1,\sft_4, \sft_3) ~.
\ee
 Here $\tilde{\sf H}$   was  defined in~\eqref{2.21} and  computed in~\eqref{2.24}.

 \paragraph{Contact diagrams. } 
 Since the quartic vertex  in \rf{3.8}  contains derivatives, the contact  contribution   may also  be represented as a sum of the three contributions 
\be\label{3.22}
\sG^\text{cont}_{abcd} (\sft_1,\sft_2,\sft_3,\sft_4)
= G_{abcd}^\text{cont}(\sft_1,\sft_2,\sft_3,\sft_4)+ G_{acbd}^\text{cont}(\sft_1,\sft_3,\sft_2,\sft_4)
+G_{abdc}^\text{cont}(\sft_1,\sft_2,\sft_4,\sft_3)~, \quad
\ee
\beqn
  G^\text{cont}_{abcd}(\sft_1,\sft_2,\sft_3,\sft_4)  &=& 
2\tilde P_{badc }
  \begin{tikzpicture}[line width=1 pt, scale=0.6, rotate=0,baseline=-0.1cm,decoration={markings, mark=at position 0.53 with {\arrow{>}}}]
\coordinate (A1) at (135:2);  \coordinate (A2) at (45:2);  
\coordinate (A3) at (-45:2);   \coordinate (A4) at (-135:2);
\coordinate (B1) at (  0, 0); \coordinate (B2) at (0,-1);
\draw[densely dashed] (0,0) circle (2);
\draw  (A1)--(B1); \draw   (B1)--(A4); 
\draw   (A3)--(B1); \draw  (B1)--(A2); 
\draw[fill=black] (A1) circle (0.1); \draw[fill=black] (A2) circle (0.1); 
\draw[fill=black] (A3) circle (0.1); \draw[fill=black] (A4) circle (0.1); 
\node[left] at (135:0.42) {$\p $};
\node[right] at ( -45:0.42) {$ \bar\p $};
\node[left] at (A1) {$\sX^a(\sft_1)$};
\node[right] at (A2) {$\sX^b(\sft_2)$};
\node[right] at (A3) {$ \sX^c(\sft_3)$};
\node[left] at (A4) {$\sX^d(\sft_4)$};
\node[above] at (B1) {$w$};
\end{tikzpicture} 
+
2\tilde P_{bcda }
  \begin{tikzpicture}[line width=1 pt, scale=0.6, rotate=0,baseline=-0.1cm,decoration={markings, mark=at position 0.53 with {\arrow{>}}}]
\coordinate (A1) at (135:2);  \coordinate (A2) at (45:2);  
\coordinate (A3) at (-45:2);   \coordinate (A4) at (-135:2);
\coordinate (B1) at (  0, 0); \coordinate (B2) at (0,-1);
\draw[densely dashed] (0,0) circle (2);
\draw  (A1)--(B1); \draw   (B1)--(A4); 
\draw   (A3)--(B1); \draw  (B1)--(A2); 
\draw[fill=black] (A1) circle (0.1); \draw[fill=black] (A2) circle (0.1); 
\draw[fill=black] (A3) circle (0.1); \draw[fill=black] (A4) circle (0.1); 
\node[left] at (135:0.42) {$\bar\p $};
\node[right] at ( -45:0.42) {$ \p $};
\node[left] at (A1) {$\sX^a(\sft_1)$};
\node[right] at (A2) {$\sX^b(\sft_2)$};
\node[right] at (A3) {$ \sX^c(\sft_3)$};
\node[left] at (A4) {$\sX^d(\sft_4)$};
\node[above] at (B1) {$w$};
\end{tikzpicture} 
\nonumber\\&&
+ 
2\tilde P_{adcb}
    \begin{tikzpicture}[line width=1 pt, scale=0.6, rotate=0,baseline=-0.1cm,decoration={markings, mark=at position 0.53 with {\arrow{>}}}]
\coordinate (A1) at (135:2);  \coordinate (A2) at (45:2);  
\coordinate (A3) at (-45:2);   \coordinate (A4) at (-135:2);
\coordinate (B1) at (  0, 0); \coordinate (B2) at (0,-1);
\draw[densely dashed] (0,0) circle (2);
\draw  (A1)--(B1); \draw   (B1)--(A4); 
\draw   (A3)--(B1); \draw  (B1)--(A2); 
\draw[fill=black] (A1) circle (0.1); \draw[fill=black] (A2) circle (0.1); 
\draw[fill=black] (A3) circle (0.1); \draw[fill=black] (A4) circle (0.1); 
\node[left] at (-135:0.42) {$\p $};
\node[right] at ( 45:0.42) {$\bar \p $};
\node[left] at (A1) {$\sX^a(\sft_1)$};
\node[right] at (A2) {$\sX^b(\sft_2)$};
\node[right] at (A3) {$ \sX^c(\sft_3)$};
\node[left] at (A4) {$\sX^d(\sft_4)$};
\node[above] at (B1) {$w$};
\end{tikzpicture} 
+ 
2\tilde P_{abcd}
    \begin{tikzpicture}[line width=1 pt, scale=0.6, rotate=0,baseline=-0.1cm,decoration={markings, mark=at position 0.53 with {\arrow{>}}}]
\coordinate (A1) at (135:2);  \coordinate (A2) at (45:2);  
\coordinate (A3) at (-45:2);   \coordinate (A4) at (-135:2);
\coordinate (B1) at (  0, 0); \coordinate (B2) at (0,-1);
\draw[densely dashed] (0,0) circle (2);
\draw  (A1)--(B1); \draw   (B1)--(A4); 
\draw   (A3)--(B1); \draw  (B1)--(A2); 
\draw[fill=black] (A1) circle (0.1); \draw[fill=black] (A2) circle (0.1); 
\draw[fill=black] (A3) circle (0.1); \draw[fill=black] (A4) circle (0.1); 
\node[left] at (-135:0.42) {$\bar\p $};
\node[right] at ( 45:0.42) {$ \p $};
\node[left] at (A1) {$\sX^a(\sft_1)$};
\node[right] at (A2) {$\sX^b(\sft_2)$};
\node[right] at (A3) {$ \sX^c(\sft_3)$};
\node[left] at (A4) {$\sX^d(\sft_4)$};
\node[above] at (B1) {$w$};
\end{tikzpicture}\hskip -.5cm . \la{3.23}
\eeqn
Here we have indicated explicitly  the coupling  tensors appearing  from each diagram 
 (the factor of 2 arises from  two ways of contracting the two legs without derivative).
 Explicitly,  we get 
\begin{align}
G^{\text{cont}}_{abcd}(\sft_1,\sft_2,\sft_3,\sft_4)
&=
 2\tilde P_{abcd }    
\Big[  \int \rmd^2 w\, \p g(\sft_1, w)  g(\sft_2, w)  \bar\p  g(\sft_3, w) g(\sft_4, w) + (\sft_1\leftrightarrow \sft_3) \Big]
\notag \\ & 
+2\tilde P_{abcd } 
\Big[  \int \rmd^2 w\,  g(\sft_1, w)   \p  g(\sft_2, w)   g(\sft_3, w) \bar \p g(\sft_4, w) + (\sft_2\leftrightarrow \sft_4) \Big]~,\la{3.24}
\end{align}
where we used 
~\eqref{3.7}. 
As a result, 
 \beqn\label{3.25}
G^{\text{cont}}_{abcd}(\sft_1,\sft_2,\sft_3,\sft_4)
=
4 \tilde P_{abcd } \; {\sf I} (\sft_1,\sft_2,\sft_3,\sft_4)~\ , 
\eeqn
where 
\begin{align}
\la{3.26}
{\sf I} (\sft_1,\sft_2,\sft_3,\sft_4) &\equiv  \int \rmd^2 w\, \p g(\sft_1, w)  g(\sft_2, w)  \bar\p  g(\sft_3, w) g(\sft_4, w) + (\sft_1\leftrightarrow \sft_3  )
\notag \\&=
  \int \rmd^2 w\,  g(\sft_1, w)   \p  g(\sft_2, w)   g(\sft_3, w) \bar \p g(\sft_4, w) +  (\sft_2\leftrightarrow \sft_4)  
  \notag \\&=
  \frac{ 1 }{\sft_{14}^2 \sft_{23}^2}\log \big( \frac{\sft_{12}\sft_{34}}{\sft_{13}\sft_{24}} \big)^2
  +  \frac{ 1 }{\sft_{12}^2 \sft_{34}^2}\log \big( \frac{\sft_{14}\sft_{23}}{\sft_{13}\sft_{24}} \big)^2
+\frac{2}{\sft_{12}\sft_{14}\sft_{23}\sft_{34}}~.
\end{align}
Collecting the contributions in \eqref{3.17}, \eqref{3.20}   and  \eqref{3.25}, \eqref{3.26}, the four-point   function in \eqref{3.16}   may be written as  
\beqn \label{3.27}
  \EV{\sfX_a(\sft_1)\sfX_b(\sft_2)\sfX_c(\sft_3)\sfX_d(\sft_4)}
  & =&\;\;
  4 \pp_s  {\sf I} (\sft_1,\sft_2,\sft_3,\sft_4)
  +4\pp_t {\sf I} (\sft_1,\sft_3,\sft_2,\sft_4)
  +4 \pp_u {\sf I} (\sft_1,\sft_2,\sft_4,\sft_3)
\nonumber\\& &+
  9 \qq_s {\sf H} (\sft_1,\sft_2,\sft_3,\sft_4)
  +9 \qq_t  {\sf H} (\sft_1,\sft_3,\sft_2,\sft_4)
  + 9\qq_u  {\sf H} (\sft_1,\sft_4,\sft_3,\sft_2)~,
  \qquad\qquad
 \eeqn
 where   in the r.h.s.  we suppressed  the indices $(a,b,c,d)$ introducing the symbolic notation 
  ($s, t, u$ stand for different channels)\foot{\ Note that the permutations of legs  on the first and second lines  of \eqref{3.27} are different.}
  \begin{align}
 \pp_s =&\tilde P_{abcd}~,     &\pp_t&=\tilde P_{acbd} ~, & \pp_u &=\tilde P_{abdc} ~,\la{3.28}
\\
 \qq_s =& Q_{ abe } Q_{ cde } ~,  & \qq_t&=Q_{ ace } Q_{ bde }~, & \qq_u&=Q_{ ade } Q_{ cbe } ~.\la{3.29}
 \end{align}
Using the   expressions  for the integrals  $\sf H$ in  (\ref{3.21}),(\ref{2.24})  and $\sf I$  in  (\ref{3.26})  
 one can compute  the four-point   function  \rf{3.27} explicitly. 
 
 We find that  the logarithmic  terms  in   \rf{3.27}  cancel  if   the following relations are satisfied  
  \be\label{3.30}
 \qq_s=\tfrac{8}{9}(\pp_s-\pp_u)~, \qquad 
 \qq_t=\tfrac{8}{9}(\pp_t-\pp_u)~, \qquad 
 \qq_u=\tfrac{8}{9}(\pp_s-\pp_t)~. \qquad 
 \ee
This implies that 
 \be\la{3.31}
 \qq_s-\qq_t-\qq_u= Q_{ abe } Q_{ cde }  -Q_{ ace } Q_{ bde }- Q_{ ade } Q_{ cbe } 
 =Q_{ abe } Q_{ cde }  +Q_{ cae } Q_{ bde } + Q_{ ade } Q_{ bce } 
  =0~,
 \ee
and also  that 
 \be\la{3.32}
\tfrac94 Q_{ abe } Q_{ cde } =\tilde P_{abcd}- \tilde P_{abdc}
= P_{abcd}+ P_{cbad}-  P_{abdc}-  P_{dbac}
=3P_{abcd} ~, \ \ \  {\rm i.e.} \ \   P_{abcd} =\tfrac34 Q_{ abe } Q_{ cde },
 \ee
 where we used  \rf{3.6} and symmetry properties  
 of  the  curvature  tensor 
 in \eqref{3.4} and \eqref{3.5}.  Written in terms of $R$  and $H$ in \rf{3.3}   this reads 
 \be\label{3.33}   R_{abcd} =\tfrac14 H_{ abe } H_{ cde }\ .  \ee
 Interestingly, the trace  of this relation, i.e. 
 $R_{ac} =\tfrac14 H_{ abe } H_{ cbe }$,   is the same as the   vanishing of the one-loop   beta-function \ci{Braaten:1985is}  of the \sm in \rf{3.1}.

  In the  group space case   \rf{3.10}  the condition \rf{3.31}    is automatically satisfied  due  to the Jacobi identity  for the structure constants.
  The condition \rf{3.32} or \rf{3.33}  reduces to 
      \be\label{3.34}
{\sf q}^2 =\tfrac43 {\sf p}~, \ \ \qquad   {\rm  i.e.} \ \ \qquad     q^2 =1 \  , 
\ee
  i.e.   is valid   only in the  WZW  model case  (cf. \rf{21},\rf{3.10}).

  \subsection{WZW  model case:    matching with  correlators of   chiral  currents}

  Thus the cancellation of the logarithmic terms in the four-point boundary  correlators  of a  generic \sm in \adst  happens only in the WZW  model. 
 This generalizes  the observation made  in section~\ref{sec2} in the $SL(2,\mathbb R)$  WZW case.
 Then the  resulting expression for the connected   four-point correlator  \rf{3.27} 
  may be written as  (using \rf{3.10}  with $q=1$, i.e. $\l^2={2 \ov k}$)
 \be\label{3.35}
 \EV{\sfX_a(\sft_1)\sfX_b(\sft_2)\sfX_c(\sft_3)\sfX_d(\sft_4)}=
 {4 \ov k}    \Big( \frac{ f_{ abe } f_{ cde } }{\sft_{12} \sft_{13}   \sft_{23}  \sft_{34}}
 +
 \frac{  f_{ ace } f_{ dbe } }{\sft_{12} \sft_{13} \sft_{23} \sft_{24}  }
 +
 \frac{  f_{ ade } f_{ bce }    }{\sft_{12} \sft_{13} \sft_{14} \sft_{23}  } \Big)~.
\ee
As in the $SL(2,\mathbb R)$ case  (cf. \rf{2.25}, \rf{2.35}), we can now explicitly check the correspondence between 
\adst boundary correlators and holomorphic correlation functions of  Kac-Moody currents. 

The basic OPE  relation  for the WZW  chiral currents on the plane  \rf{1.3}   gives the two-point function \rf{1.4}. 
Higher point correlators  can  be  obtained by repeatedly using the OPE (\ref{1.3}).\foot{\ 
The current-current OPE translates into the (recursion) relation  \cite{Zamolodchikov:1985wn}
\be 
 \EV{J_{a_1}(w_1)...J_{a_n}(w_n) }=\sum_{j=2}^n \EV{J_{a_2}(w_2)...J_{a_ {j-1}}(w_{j-1})
 \Big[ \frac{k\delta_{a_1 a_{j}}}{(w_1-w_j)^2} + \frac{ f_{a_1 a_j c } J_c(w_j)}{w_1-w_j} \Big]\, 
 J_{a_ {j+1}}(w_{j+1}) ...J_{a_n}(w_n)  }~.\no 
 \ee
The  mutual locality of the KM currents implies a trivial (meromorphic) singularity structure 
and the solution of the above relation is simply obtained by isolating poles as in
 (\ref{3.36})-(\ref{3.38}).
 }
In particular, one finds 
\beqn\label{3.36}
\EV{J_a(w_1) J_b(w_2) J_c(w_3)} &=&\frac{k f_{abc}}{w_{12}w_{13}w_{23}}~,
\\
\EV{J_a(w_1) J_b(w_2) J_c(w_3) J_d(w_4) } 
&=&\frac{k^2 \delta_{ab} \delta_{cd} }{ w_{12}^2 w_{34}^2}
+\frac{k f_{abe}f_{cde}}{ w_{12}w_{34}w_{23}w_{24} }
+\PBK{ w_2\leftrightarrow w_3 \\ b \leftrightarrow c}
+\PBK{ w_2\leftrightarrow w_4 \\ b \leftrightarrow d}~.\la{3.37}
\qquad
\eeqn
The ``connected''    part of \rf{3.37} may be written as 
\beqn\no 
\EV{J_a(w_1) J_b(w_2) J_c(w_3) J_d(w_4) }_\text{conn}
&=&k \Big( 
 \frac{  f_{abe}f_{cde}}{ w_{12}w_{34}w_{23}w_{24} }
 + \frac{  f_{ace}f_{dbe}}{ w_{13}w_{24}w_{23}w_{34} }
+ \frac{  f_{ade}f_{bce}}{ w_{14}w_{ 23}w_{43}w_{24} }
\Big)
\nonumber \\& =&  
k \Big( \frac{   f_{abe}f_{cde}    }
{w_{12}w_{13} w_{23} w_{34}}
+
 \frac{     f_{ace}f_{dbe}    }
{w_{12}w_{13} w_{23}w_{24} }
+
 \frac{     f_{ade}f_{bce}    }
{w_{12}w_{13}w_{14}w_{23}  }
\Big)~.\qquad \la{3.38}
\eeqn
 Here in  the second line we wrote an equivalent  expression 
  (expressing  crossing symmetry of the four-point function)  that is a consequence of the Jacobi identity for the structure constants. 
  
  Restricting  the points to the real line ($w_i\to \sft_i$)  we can  identify the two-point \rf{1.4} and three-point \rf{1.5} correlators 
  of the  currents  with the  corresponding boundary correlators  in \rf{3.14} and \rf{3.15}   up to an overall   universal  factor  $\kappa^n$  where  $n=2,3,....$ is the number of legs.  Explicitly,  this  amounts to  the formal  identification (assuming $k >0$)
    \be\label{3.39}
 \sfX^a  \to   \kappa J^a~,\qquad \qquad    \kappa =  \sqrt{ \tfrac{2 }{ k}}   \ . 
  \ee
 Indeed, the two-point  functions match if  $\kappa^2= {2\ov k}$, while 
   the three-point  functions  match  for  
    $Q_{abc}$  in \rf{3.15}   related to $f_{abc}$ in  \rf{3.36} 
    as in \rf{3.10}  and $\kappa^3=  { 2 \ov k}  \l $. 
 Furthermore, 
  the four-point correlator \rf{3.35}   is also in precise agreement with the  boundary restriction of the connected part of the correlator of four currents in \rf{3.38}.  
  
  As  already  mentioned in the Introduction,  one can   give a simple semiclassical argument  supporting the relation \rf{3.39}, i.e.
  the expression for $\kappa$ that, we remark, is same as in (\ref{2.39}).
  Starting with the expression for the   $w$-component of the chiral  current 
    consistent with $k >0$ and  \rf{1.3},\rf{3.36} (see, e.g.,   \ci{DiFrancesco:1997nk})
  $J^a =  - k\tr ( t^a  \p  g \, g^{-1})$
  and  using the parametrization  $g=e^{-i \lambda t_a X^a}$  
     we get  in the $\sfz \to 0$ limit (for the  boundary asymptotics in \rf{3.11}) 
\be\la{3.40}
\sfz \to 0: \ \ \ \ \  J^a= i k \lambda  \p X^a +\cdots = i k \lambda   \, \tfrac12 ( \p_\sft-i \p_\sfz) (\sfz \sfX^a+\cdots) 
= \ha k \lambda  \sfX^a  +\cdots~.
\ee
This   suggests the identification 
$J^a= \ha k \lambda   
\sfX^a $ as in \rf{3.39}  where $\l$   is  given by  \rf{21}, i.e. 
 $\kappa =\sqrt{2\ov  k}$.

\section{Quantum   corrections to boundary correlators in  $SL(2,\mathbb R)$ WZW model} \la{sec4}

The above discussion was restricted to consideration of tree-level  terms in the boundary correlators in AdS$_2$.
Let us now try to  test   the relation between   the  boundary correlators of  WZW  fields  and  chiral  currents \rf{1.2} 
beyond  the   classical (large $k$)  limit.  This  requires determining    loop corrections to \adst  boundary correlators. 
Similar computations were   done in the Liouville  and Toda theories in 
 \cite{Beccaria:2019stp,Beccaria:2019mev,Beccaria:2019dju}  and it was found 
 that  the  analogs of the coefficient $\kappa$   in \rf{1.2}  that relate    boundary correlators of elementary 
 fields in \adst  to correlators of CFT currents (stress tensor and 
 $\mc W$-symmetry  currents)  restricted to real line 
 receive quantum corrections.  
 
 In the  present WZW  model  case, the simplicity of the 
semiclassical argument in  (\ref{3.40})  suggests instead that the relation \rf{1.2} or (\ref{2.39}),\rf{3.39}
  may be  exact.\foot{\ One  could   wonder if    
   the  level   $k$ in \rf{1.2}   may get a familiar  quantum   shift by the 
  dual Coxeter number  of $G$  (i.e. $k \to k + c_G$)
which is  known to appear 
from the quantum jacobian transformation 
from the group fields $g$ to currents and in the Sugawara construction of the stress tensor and related computation of the 
central charge. As we shall  argue   below, there exists a   natural computation scheme in which 
this   does not apparently happen  in the  present case of $\kappa$ in \rf{1.2}.}


To provide   support  to  this conjecture below we shall consider the computation of  one-loop corrections  to the two-point and three-point boundary correlators \rf{1.1} 
on the example of the $SL(2, \mathbb R)$  WZW model.
 A central issue    will be the choice of a   UV regularization  and  subtraction  scheme    consistent with underlying 
 $SL(2, \mathbb R)$  symmetry of the model.  
 It turns  out to be possible  to   relate the  scheme    ambiguity  to the  definition of 
 the  propagator at the coinciding points, i.e. to the choice of  the renormalized value of the   
self-contraction contributions.
 

\subsection{One-loop corrections to the  two-point  correlators } \la{1-loop2pt}

Let us start   with computing the one-loop corrections to the tree-level 
 two-point functions \rf{2.13}  for the fields in the action (\ref{2.5}), i.e. to 
the  boundary correlators  $\langle \sfPsi\widetilde{\sfPsi}\rangle$ and $\langle\sfPhi\sfPhi\rangle$.

\subsubsection{$\langle \sfPsi\widetilde{\sfPsi} \rangle$} \la{2ptLoopPsi}

One-loop corrections  to  the  $\psi,\tilde \psi $  propagator  in \adst    come from the following  diagrams:
\be
\la{4.1}
\begin{tikzpicture}[line width=1 pt, scale=0.8, baseline=-0.1cm,decoration={markings, mark=at position 0.53 with {\arrow{>}}}]
\coordinate (A1) at (-0.8,0);        \coordinate (A2) at (0.8,0); 
\coordinate (B1) at (-2,0);        \coordinate (B2) at (2,0); 
\draw[postaction=decorate] (B1)--(A1); 
\draw[postaction=decorate] (A1)--(A2);
\draw[postaction=decorate] (A2)--(B2);
\draw (A2) arc(0:180:0.8);
\node[below] at (0,0) {$\scriptstyle \rm regular$};
\end{tikzpicture}+\Big[
\begin{tikzpicture}[line width=1 pt, scale=0.8, baseline=-0.1cm,decoration={markings, mark=at position 0.53 with {\arrow{>}}}]
\coordinate (A1) at (-0.8,0);        \coordinate (A2) at (0.8,0); 
\coordinate (B1) at (-2,0);        \coordinate (B2) at (2,0); 
\draw[postaction=decorate] (B1)--(A1); 
\draw[postaction=decorate] (A1)--(A2);
\draw[postaction=decorate] (A2)--(B2);
\draw (A2) arc(0:180:0.8);
\node[below] at (0,0) {$\scriptstyle \delta-\rm function$};
\end{tikzpicture}+
\begin{tikzpicture}[line width=1 pt, scale=0.8, baseline=-0.1cm,decoration={markings, mark=at position 0.53 with {\arrow{>}}}]
\coordinate (A1) at (-0.8,0);        \coordinate (A2) at (0.8,0); 
\coordinate (B1) at (-2,0);        \coordinate (B2) at (2,0); 
\draw[postaction=decorate] (B1)--(0,0); 
\draw[postaction=decorate] (0,0)--(B2);
\draw (0,0.5) circle(0.5);
\end{tikzpicture}  \Big] 
+
 \begin{tikzpicture}[line width=1 pt, scale=0.5, rotate=0,baseline=-0.1cm,decoration={markings, mark=at position 0.53 with {\arrow{>}}}]
\draw [postaction={decorate}]  (-2.2,0)--(0,0);  
\draw [postaction={decorate}]  (0,0)--(2.2,0);  
\draw [postaction={decorate}]    (0.8,1.2) arc(0:180:0.8);
\draw ( -0.8 ,1.2) arc( 180:360:0.8);
\draw (0,0) --(0,0.4);
\end{tikzpicture}~. 
\ee
Here we have separated  the contributions of the  regular  and $\delta$-function terms   in   (\ref{2.11}) combining the latter with  
 the 
 self-contraction diagram   corresponding to the   vertex  $\phi^{2}\psi\psib$  in \rf{2.5}  as 
both  are proportional to   the   free scalar   propagator at the coinciding points, i.e. $g(w,w)$  (cf. \rf{2.9},\rf{2.10}).
The last  tadpole diagram with a $\psi$ loop is linearly divergent and  may  be removed by 
imposing the normalization condition $\langle\phi\rangle=0$.

The first  contribution  in \rf{4.1}  involving the { regular} part of the  second derivative of the  propagator in \rf{2.11} 
(with legs taken to the boundary)  is given by 
\be\la{42}
D_{\sfPsi\widetilde{\sfPsi}}(\bt_{12}) = 
\begin{tikzpicture}[line width=1 pt, scale=0.8, baseline=-0.1cm,decoration={markings, mark=at position 0.53 with {\arrow{>}}}]
\coordinate (A1) at (-0.8,0);        \coordinate (A2) at (0.8,0); 
\coordinate (B1) at (-2,0);        \coordinate (B2) at (2,0); 
\node[above] at (B1) {$\bt_{1}$};   \node[above] at (B2) {$\bt_{2}$};
\draw[postaction=decorate] (B1)--(A1); 
\draw[postaction=decorate] (A1)--(A2);
\draw[postaction=decorate] (A2)--(B2);
\draw (A2) arc(0:180:0.8);
\node[below] at (0,0) {$\scriptstyle \rm regular$};
\end{tikzpicture} = 
2^{2}\, \frac{b^2}{\pi^2}\,\widehat E(\bt_{12})~,
\ee
\begin{align}
\widehat{E}(\bt_{12}) &\equiv  \int d^2w\ d^2w' \ \frac{1}{(\bt-\bt_{1}-i\bz)^{2}}\frac{1}{(\bt'-\bt_{2}+i\bz')^{2}}
\frac{1}{[\bt-\bt'+i\,(\bz+\bz')]^{2}}\,g(w,w')\notag \\
&= \int d^2w\ d^2w' \ \frac{1}{(\wb-\bt_{1})^{2}}\frac{1}{(w'-\bt_{2})^{2}}
\frac{1}{(w-\wb')^{2}}\,g(w,w')    = \p_{\bt_{1}}\p_{\bt_{2}}  J (\bt_{12}) \ , \la{4.3}   \\
J(\bt_{12}) & \equiv \int d^2w\ d^2w' \ \frac{1}{\wb-\bt_{1}}\frac{1}{w'-\bt_{2}}
\frac{1}{(w-\wb')^{2}}\,g(w,w') \ . \la{433}
\end{align}
By  formal shifting  and rescaling $w, w'$ one may try to  argue   that 
 the  integral $J$     should be   independent of $\bt_{1},\bt_{2}$.  
 However, it is IR   divergent and thus requires a regularization.
A regularization  will   then be  expected to  give $J \sim \log ( \Lambda^{-1}   |\bt_{1}-\bt_{2}|)$   and thus   a finite $\sim { 1\ov \bt_{12}^{2}}$
contribution to $\widehat{E}(\bt_{12})$. 
Indeed, integrating by parts  the formal expression in \rf{433}  we get  (using \rf{2.10})\foot{\ The integral 
 over $\sfz, \sfz'$  here 
may be split and turned into  a double integral over $z, Z$ with 
$0<z<Z$. Then setting 
$
Z = y \sft_{12}\ , \ 
z =  x y \sft_{12} $  one is to integrate over 
$ 0<x<1$ and $0<y<\infty$.}
\be
 J   
= \int d^2w d^2w' \ \frac{1}{\wb-\bt_{1}}\frac{1}{w'-\bt_{2}}
\frac{1}{w-\wb'}\,\p_{w}g(w,w')= \int_{0}^{1}dx\int_{0}^{\infty}dy\ 
\frac{\pi^{2}(1+6\,i\,y+2\,i\,x\,y-8\,y^{2})}{(-i+2y)(-i+2y+2xy)^{2}}~.
\ee
This  is divergent due to the contribution from the $y\to +\infty$ region where
 the integrand  scales as 
  $\sim 1/(x^{2}y)$. 
A cutoff on the $\bz, \bz'$  integrals near zero 
 in \rf{433}  translates into the modified integration range $0<y<\frac{\Lambda}{\bt_{12}}, \ \Lambda \to \infty$.
Then we find for the regularized integral
\footnote{\ We first integrate over $x$ and then add and subtract the leading term of the $y\to \infty$
expansion.}
\begin{align} \la{45}
J(\bt_{12}; \Lambda) =\int_{0}^{1}dx\int_{0}^{\Lambda/\bt_{12}}dy\ 
\frac{\pi^{2}(1+6\,i\,y-2\,i\,x\,y+8\,y^{2})}{(-i+2y)(-i+2y+2xy)^{2}} \stackrel{\Lambda\to\infty}{=}
-\frac{\pi^{2}}{2} \log{\Lambda\ov \bt_{12} }+\text{finite}~,
\end{align}
 and thus  
\be
\la{4.6}
\widehat{E}(\bt_{12}) = \p_{\bt_{1}}\p_{\bt_{2}}\Big[-\frac{\pi^{2}}{2}\log{\Lambda\ov \bt_{12}}+\cdots\Big]
= \frac{\pi^{2}}{2\,\bt_{12}^{2}}~.
\ee
Including also the  contribution of the 
 square bracket  terms in \eqref{4.1}  which depend on  regularized value of $g(w,w)$  we finish with 
 the following   one-loop  (i.e.  order $b^2 \sim {1\ov k}$) 
 correction  to the  tree-level  boundary correlator   \rf{2.13}
\be
\la{4.10}
\langle\sfPsi(\bt_{1})\widetilde{\sfPsi}(\bt_{2})\rangle_{\rm 1-loop} 
 = \frac{2 b^2 }{\bt_{12}^{2}}\, (1- g_0)  \ , \qquad \qquad g_0 \equiv g(w,w) \ . 
\ee
Thus a particular   scheme choice where $g_0=1$    would  lead to   the vanishing of the  one-loop  
correction. 

To put this in a  more general   context, while the WZW  is UV finite in the sense  that there is no coupling renormalization, 
 there may still be a wave 
 function renormalization (i.e. UV divergent $Z$-factor in the off-shell 2-point function). This 
   should be accounted  for in the definition of the S-matrix: 
   the scattering amplitudes  defined    in terms of  correlators with extra 
  powers of $Z$  will be automatically finite (see, e.g., a discussion in \ci{Roiban:2014cia} and refs. there). 
  Similar considerations   should apply to  the analog of  S-matrix in AdS (see section 5)  
  and thus   to the boundary  correlators. 
  Here we  will   effectively  
  by-pass this subtlety   by  simply assuming a particular subtraction 
   under which the wave-function renormalization factor is trivial.\foot{For some recent discussions 
   of  one-loop  self-energy corrections  in AdS  see  \ci{Giombi:2017hpr,Bertan:2018afl,Meltzer:2019nbs}.}

\subsubsection{$\langle\sfPhi\sfPhi\rangle$}

The one-loop correction to  the  boundary two-point   function 
$\langle\sfPhi(\bt_{1})\sfPhi(\bt_{2})\rangle$  is given by the  sum of  two diagrams:  a bubble and a 
self-contraction diagram.

\paragraph{Bubble.} The bubble contribution is 
\be
\la{4.8}
\begin{tikzpicture}[line width=1 pt, scale=0.8, baseline=-0.1cm,decoration={markings, mark=at position 0.53 with {\arrow{>}}}]
\coordinate (A1) at (-0.8,0);        \coordinate (A2) at (0.8,0); 
\coordinate (B1) at (-2,0);        \coordinate (B2) at (2,0); 
\node[above] at (B1) {$\bt_{1}$};   \node[above] at (B2) {$\bt_{2}$};
\draw (A1)--(B1); \draw (A2)--(B2);
\draw[postaction=decorate] (A2) arc(0:180:0.8);
\draw[postaction=decorate] (A1) arc(-180:0:0.8);
\end{tikzpicture} = 
4\frac{b^2}{\pi^2}\Big[\widehat D(\bt_{12}) +\pi\,\widehat D_{+}(\bt_{12})
+\pi\,\widehat D_{-}(\bt_{12})+\pi^2\,\widehat D_{\rm cont}(\bt_{12}) \Big]~.
\ee
Here we decomposed  the  derivatives of  both  propagators \rf{2.11} in the loop  into the regular and $\delta$-function parts  getting  four terms:
with no $\delta$-function  factors ($\widehat D$), with one ($\widehat D_{\pm}$)  and with two ($\widehat D_{\rm cont}$). 
Explicitly,  
\begin{align}
\widehat{D}(\bt_{12}) &=
 \int d^{2}w\ d^{2}w' \ \frac{\bz}{(\bt-\bt_{1})^{2}+\bz^{2}}\ 
 \frac{\bz'}{(\bt'-\bt_{2})^{2}+\bz'^{2}}\frac{1}{[(\bt-\bt')^{2}+(\bz+\bz')^{2}]^{2}}~,\notag \\
\widehat{D}_{\pm}(\bt_{12}) &= \int d^{2}w \ d^{2}w'\ \frac{\bz}{(\bt-\bt_{1})^{2}+\bz^{2}}\ 
 \frac{\bz'}{(\bt'-\bt_{2})^{2}+\bz'^{2}}\frac{1}{[\bt-\bt'\pm i(\bz+\bz')]^{2}}\,\delta(w-w')\notag \\
 &=  -\frac{1}{4}\int d^{2}w \ \frac{1}{(\bt-\bt_{1})^{2}+\bz^{2}}\ 
 \frac{1}{(\bt-\bt_{2})^{2}+\bz^{2}}~.\la{410}
\end{align}
Integrating over $\bt, \bt', \bz'$  gives 
\be\la{4.11}
\widehat D(\bt_{12}) = \frac{\pi^{2}}{2}\,\int_{0}^{\infty}d\bz\frac{1}{\bz\,(\bt_{12}^{2}+4\bz^{2})}~,\qquad
\widehat D_{\pm}(\bt_{12}) = -\frac{\pi}{2}\,\int_{0}^{\infty}d\bz\frac{1}{\bz\,(\bt_{12}^{2}+4\bz^{2})}
= -\frac{1}{\pi}\widehat D(\bt_{12})~.
\ee

\paragraph{Self-contraction.} 
With the same  decomposition of the   two derivatives of the propagator in the loop  \rf{2.11}  we get 
\be
\begin{tikzpicture}[line width=1 pt, scale=0.8, baseline=-0.1cm,decoration={markings, mark=at position 0.53 with {\arrow{>}}}]
\coordinate (A1) at (-0.8,0);        \coordinate (A2) at (0.8,0); 
\coordinate (B1) at (-2,0);        \coordinate (B2) at (2,0); 
\node[above] at (B1) {$\bt_{1}$};   \node[above] at (B2) {$\bt_{2}$};
\draw (B1)--(B2);
\draw[postaction=decorate] (0,0) arc(-90:270:0.8);
\end{tikzpicture} =-
4\frac{b^{2}}{\pi}\   \Big[\widehat  D_{\pm}(\bt_{12})  +\pi\,\widehat D_{\rm cont}(\bt_{12}) \Big]\ .\la{4.12}
\ee
As a result, 
 $\widehat D_{\rm cont}$  here   exactly cancels  against   the double $\delta$-function part  in the  bubble  diagram \eqref{4.8}.
The total expression for the one-loop  correction is then 
\be
\la{4.15}
\langle\sfPhi\sfPhi\rangle_{\rm 1-loop}=
\frac{4b^{2}}{\pi^{2}}\big(\widehat D+2\pi\widehat D_{\pm}\big)-\frac{4b^{2}}{\pi}\widehat D_{\pm}=
\frac{4b^{2}}{\pi^{2}}\big(\widehat D+\pi\widehat D_{\pm}\big)=0~,
\ee
where we used \rf{4.11}, i.e. $\widehat D_{\pm}=-\frac{1}{\pi}\widehat D$.

A   more rigorous  derivation of \rf{4.15}  requires introducing a regularization   factor $\bz^{\eps}$ in each 
(formally divergent)  AdS integral. Then
\begin{align}
\widehat{D}_{\eps}(\bt_{12}) &= \int d^{2}w\ d^{2}w' \ \bz^{\eps}\bz'^{\eps}\frac{\bz}{(\bt-\bt_{1})^{2}+\bz^{2}}\ 
 \frac{\bz'}{(\bt'-\bt_{2})^{2}+\bz'^{2}}\frac{1}{[(\bt-\bt')^{2}+(\bz+\bz')^{2}]^{2}} \notag \\
 &=\frac{1}{\bt_{12}^{2-2\eps}}\frac{\pi ^3 4^{-\eps } (2 \eps +1) \cot (\pi  \eps ) \Gamma (-2
   \eps -2 ) \Gamma (\eps +2)}{\Gamma (-\eps )}~, \notag \\
 \widehat{D}_{\pm, \eps}(\bt_{12}) &=  -\frac{1}{4}\int\ d^{2}w\ \bz^{2\eps} \ \frac{1}{(\bt-\bt_{1})^{2}+\bz^{2}}\ 
 \frac{1}{(\bt-\bt_{2})^{2}+\bz^{2}} = -\frac{1}{\bt_{12}^{2-2\eps}}\,2^{-2 -2 \eps}\pi^{2}\,\csc(\pi\eps)\ . 
 \end{align}
 Then expanding for  small $\eps$ gives 
 \be
 \widehat{D}_{\eps}(\bt_{12})  =   \frac{1}{\bt_{12}^{2}}\,\Big[
   \frac{\pi^{2}}{4\eps}+\frac{\pi^{2}}{4}\log\frac{\bt_{12}^{2}}{4}+\cdots\Big]~,\qquad
    \widehat{D}_{\pm, \eps}(\bt_{12}) =  \frac{1}{\bt_{12}^{2}}\,\Big[
   -\frac{\pi}{4\eps}-\frac{\pi}{4}\log\frac{\bt_{12}^{2}}{4}+\cdots\Big]\ , 
 \ee
 leading again to    \rf{4.15}. 

 Compared to \rf{4.10}  the  vanishing   result in  \rf{4.15}  suggests
that  for  consistency   with global   symmetry (see \rf{B.7})) 
 the   value  of $g_0=g(w,w)$   to be used in (\ref{4.10})   should be indeed 
\be
\la{413}
g_0 =1\ .
\ee

\subsection{One-loop correction to the  three-point  correlator}

Let us now compute  the one-loop correction to the three-point function $\langle\sfPsi(\sft_1) \widetilde{\sfPsi} (\sft_2)\sfPhi (\sft_3) \rangle $
with the tree-level value given by 
in \rf{2.17}.\footnote{\ Similar loop corrected three-point functions in \adst
have been  considered 
in the Liouville theory \cite{Beccaria:2019stp}, the abelian or non-abelian
Toda theory \cite{Beccaria:2019mev}, 
and the $\mc N=1$ supersymmetric Liouville theory \cite{Beccaria:2019dju}. In all  those cases, 
the analysis has  been semi-analytic because some contribution required a numerical 
evaluation. In the present WZW model all calculations will be fully analytical 
due to the simpler structure of virtual exchanges.
}
There are two types of contributions:    from  the triangle diagram  and its ``limits'',
and from the  ``self-energy''  corrections to  the propagators in the tree-level diagram \rf{215}.

\paragraph{Triangle. }  The first   is given by  the following set of diagrams
 \be
 \begin{tikzpicture}[line width=1 pt, scale=0.7, rotate=0,baseline=-0.1cm,decoration={markings, mark=at position 0.53 with {\arrow{>}}}]
 \coordinate (A1) at (90:2);  \coordinate (A2) at (210:2);  \coordinate (A3) at (-30:2);
\coordinate (B1) at (90:1);  \coordinate (B2) at (210:1);  \coordinate (B3) at (-30:1);
\draw[dashed ] (0,0) circle (2);
 \draw (B2)--(B3) ;
\draw [postaction={decorate}]  (B2)--(B1);
\draw [postaction={decorate}]  (B1)--(B3);
\draw (A1)--(B1); \draw [postaction={decorate}]  (A2)--(B2); \draw [postaction={decorate}]  (B3)--(A3);
\draw[fill=black] (A1) circle (0.07); \draw[fill=black] (A2) circle (0.07); \draw[fill=black] (A3) circle (0.07); 
\node[left] at (A2) {$\sft_1$};
\node[right] at (A3) {$\sft_2$};
\node[above] at (A1) {$\sft_3$};
\node[above] at (B2) {$w'\quad $};
\node[above] at (B3) {$\quad w''$};
\node[right] at (B1) {$w  $};
 \node[below]  at  (0,-2.3) {$ (a) $};
\end{tikzpicture}
+
 \begin{tikzpicture}[line width=1 pt, scale=0.7, rotate=0,baseline=-0.1cm,decoration={markings, mark=at position 0.53 with {\arrow{>}}}]
\coordinate (A1) at (90:2);  \coordinate (A2) at (210:2);  \coordinate (A3) at (-30:2);
\coordinate (B1) at (90:1);  \coordinate (B2) at (210:1);  \coordinate (B3) at (-30:1);
\draw[dashed ] (0,0) circle (2);
\draw  (A1)--(0,0); 
\draw [postaction={decorate}] (A2)--(0,0); 
\draw  [postaction={decorate}]   (-30   : 1.4  )--(A3);
\draw[postaction={decorate}]  (-30   : 0  )arc(   150:-30:0.7);
\draw    (-30   : 1.4  )arc(  330:150:0.7);
\draw[fill=black] (A1) circle (0.07); \draw[fill=black] (A2) circle (0.07); \draw[fill=black] (A3) circle (0.07); 
\node[left] at (A2) {$\sft_1$};
\node[right] at (A3) {$\sft_2$};
\node[above] at (A1) {$\sft_3$};
\node[left] at (0,0) {$w \; $};
\node[left] at (-30   : 1.4 ) {$w''$};
 \node[below]  at  (0,-2.3) {$ (b) $};
\end{tikzpicture}
  + 
 \begin{tikzpicture}[line width=1 pt, scale=0.7, rotate=0,baseline=-0.1cm,decoration={markings, mark=at position 0.53 with {\arrow{>}}}]
\coordinate (A1) at (90:2);  \coordinate (A2) at (210:2);  \coordinate (A3) at (-30:2);
\coordinate (B1) at (90:1);  \coordinate (B2) at (210:1);  \coordinate (B3) at (-30:1);
\draw[dashed ] (0,0) circle (2);
\draw  (A1)--(0,0); 
\draw [postaction={decorate}] (0,0)--(A3); 
\draw  [postaction={decorate}]   (A2)--(210   : 1.4  ) ;
\draw[postaction={decorate}]  (210   : 1.4  )arc(   210:30:0.7);
\draw    (210   : 1.4  )arc(   210:390:0.7);
\draw[fill=black] (A1) circle (0.07); \draw[fill=black] (A2) circle (0.07); \draw[fill=black] (A3) circle (0.07); 
\node[left] at (A2) {$\sft_1$};
\node[right] at (A3) {$\sft_2$};
\node[above] at (A1) {$\sft_3$};
\node[right] at (B2) {$\!\!\!\!w '$};
\node[right] at (0,0) {$\,w $};
 \node[below]  at  (0,-2.3) {$ (c) $};
\end{tikzpicture}
+
  \begin{tikzpicture}[line width=1 pt, scale=0.7, rotate=0,baseline=-0.1cm,decoration={markings, mark=at position 0.53 with {\arrow{>}}}]
\coordinate (A1) at (90:2);  \coordinate (A2) at (210:2);  \coordinate (A3) at (-30:2);
\coordinate (B1) at (90:1);  \coordinate (B2) at (240:1);  \coordinate (B3) at (-60:1);
\draw[dashed ] (0,0) circle (2);
\draw  (A1)--(0,0); 
\draw [postaction={decorate}] (0,0)--(A3); 
\draw  [postaction={decorate}]   (A2)--(0,0  ) ;
\draw plot [smooth, tension=1] coordinates {(0,0) (234:0.955) (270:1.5) (-54:0.95) (0,0)};
\draw[fill=black] (A1) circle (0.07); \draw[fill=black] (A2) circle (0.07); \draw[fill=black] (A3) circle (0.07); 
\node[left] at (A2) {$\sft_1$};
\node[right] at (A3) {$\sft_2$};
\node[above] at (A1) {$\sft_3$};
\node[right] at (0,0) {$w $};
 \node[below]  at  (0,-2.3) {$ (d) $};
 \end{tikzpicture}\la{417}
 \ee
 i.e. 
%
 \be\la{418}
 \langle\sfPsi(\sft_1) \widetilde{\sfPsi} (\sft_2)\sfPhi(\sft_3) \rangle^{\text{triangle}}_{\text{1-loop}}
 = 
 V_{a}(\bt_{1}, \bt_{2}, \bt_{3}) +V_{b}(\bt_{1}, \bt_{2}, \bt_{3}) +V_{c}(\bt_{1}, \bt_{2}, \bt_{3}) +V_{d}(\bt_{1}, \bt_{2}, \bt_{3}).
 \ee 
 We will again  separate the contributions coming from regular and   $\delta$-function parts  of   derivatives of  internal  propagators in  \rf{2.11}.
 Then the ``regular'' part of $V_a$  is given 
 by~\footnote{\ In the integration by parts
 we  may ignore the $\delta$-function       from $\p_{w'}\frac{1}{\wb'-\bt_{1}}$ and its derivatives as they localize the bulk integral to the boundary. }  
\begin{align}
 V^{\text{reg}}_{a} 
&= \Big( \frac{-2b}{\pi} \Big)^{3}  \int \!\! d^{2}w\,d^{2}w'\,d^{2}w''\,\frac{1}{(\wb'-\bt_{1})^{2}}\frac{1}{(w'-\wb)^{2}}\frac{1}{(w-\wb'')^{2}}
\frac{1}{(w''-\bt_{2})^{2}}\frac{\bz}{(\bt-\bt_{3})^{2}+\bz^{2}}\,g(w',w'')\notag \\
&=\Big( \frac{-2b}{\pi} \Big)^{3}  \int \!\!  d^{2}w\,d^{2}w'\,d^{2}w''\,\frac{1}{(\wb'-\bt_{1})^{2}}\frac{\p_{w'}g(w',w'')}{w'-\wb}\frac{1}{(w-\wb'')^{2}}
\frac{1}{(w''-\bt_{2})^{2}}\frac{\bz}{(\bt-\bt_{3})^{2}+\bz^{2}}~,
\end{align}
This is the triple integral over a half-plane of  a rational  integrand.
Applying the residue theorem gives 
\be\la{4.20}
 V^{\text{reg}}_{a}(\bt_{1}, \bt_{2}, \bt_{3}) =\Big( \frac{-2b}{\pi} \Big)^{3}     \frac{\pi^{3}}{4}\,\frac{i}{\bt_{12}\bt_{13}\bt_{23}}
 = -\frac{2ib^3 }{\bt_{12}\bt_{13}\bt_{23}}~.
\ee
The  contribution  $V^{\delta}_{a}$   with only  one $\delta$-function from  \rf{2.11} 
turns out to precisely cancel  the regular parts of the    contributions of the  two diagrams  $(b)$ and $(c)$, i.e. 
$V^{\text{reg}}_{b}+V^{\text{reg}}_{c}$. 

The contributions of the  diagrams with the $\delta$-function  parts   of the derivatives of all of the  internal  $(\psi,\tilde \psi)$   propagators  
  reduce  to that of the  diagram $(d)$  (with different overall factors). 
 Explicitly, the contribution 
    $V^{2 \delta}_{a}$   with the two $\delta$-functions $ \delta^{(2)}(w-w'),\ \delta^{(2)}(w-w'')$  in   the diagram $(a)$  
    is given by   $V^{2 \delta}_{a}  = - 2 V_d$ (accounting for the symmetry factor  of the $\phi$ loop) and 
$
 V^{1 \delta}_{b} =V^{1 \delta}_{c}=-V^{2 \delta}_{a}$, i.e.  
  \be\la{4.21} 
   V^{ \delta}(\bt_{1}, \bt_{2}, \bt_{3}) =V^{2 \delta}_{a}+ V^{1 \delta}_{b} +V^{1 \delta}_{c}+ V_{d} =\tfrac12 V^{2 \delta}_{a}
       = \frac{2 ib^3 g(w,w) }{\sft_{12}\sft_{23}\sft_{13}}~.\qquad
     \ee
 Thus finally (using the notation $g_0$  in \rf{4.10})  
 \be\la{4.22}
  \langle\sfPsi(\sft_1) \widetilde{\sfPsi} (\sft_2)\sfPhi(\sft_3) \rangle^{\text{triangle}}_{\text{1-loop}}
  =- \frac{2ib^3 }{\sft_{12}\sft_{23}\sft_{13}}   (1-   g_0) ~ .
 \ee

\paragraph{Self-energy corrections. }  
The contribution of the corresponding diagrams
 (here  gray circles  stand  for sums of  relevant one-loop diagrams as in 
 \rf{4.1} and \rf{4.8},\rf{4.12})
 \be
\begin{tikzpicture}[line width=1 pt, scale=0.7, rotate=0,baseline=-0.1cm,decoration={markings, mark=at position 0.53 with {\arrow{>}}}]
\coordinate (A1) at (90:2);  \coordinate (A2) at (210:2);  \coordinate (A3) at (-30:2);
\coordinate (B1) at (90:1);  \coordinate (B2) at (240:1);  \coordinate (B3) at (-60:1);
\draw[dashed ] (0,0) circle (2);
\draw  (A1)--(0,0); 
\draw [postaction={decorate}] (0,0)--(A3); 
\draw  [postaction={decorate}]   (A2)--(0,0  ) ;
\draw[fill=white]  (0,1) circle (0.5);
\draw[fill=black!10] (0,1) circle(0.5);
\draw[fill=black] (A1) circle (0.07); \draw[fill=black] (A2) circle (0.07); \draw[fill=black] (A3) circle (0.07); 
\node[left] at (A2) {$\sft_1$};
\node[right] at (A3) {$\sft_2$};
\node[above] at (A1) {$\sft_3$};
 \end{tikzpicture}
 +
   \begin{tikzpicture}[line width=1 pt, scale=0.7, rotate=0,baseline=-0.1cm,decoration={markings, mark=at position 0.7  with {\arrow{>}}}]
\coordinate (A1) at (90:2);  \coordinate (A2) at (210:2);  \coordinate (A3) at (-30:2);
\coordinate (B1) at (90:1);  \coordinate (B2) at (240:1);  \coordinate (B3) at (-60:1);
\coordinate (C1) at (-30:0.5);  \coordinate (C2) at (-30:1.5); \coordinate (C0) at (-30:1); 
\draw[dashed ] (0,0) circle (2);
\draw  (A1)--(0,0); 
\draw [postaction={decorate}] (0,0)--(C1); 
\draw [postaction={decorate}] (C2)--(A3); 
\draw  [postaction={decorate}]   (A2)--(0,0) ;
\draw[fill=white]  (C0) circle (0.5);
\draw[fill=black!10] (C0) circle(0.5);
\draw[fill=black] (A1) circle (0.07); \draw[fill=black] (A2) circle (0.07); \draw[fill=black] (A3) circle (0.07); 
\node[left] at (A2) {$\sft_1$};
\node[right] at (A3) {$\sft_2$};
\node[above] at (A1) {$\sft_3$};
 \end{tikzpicture}
 +   \begin{tikzpicture}[line width=1 pt, scale=0.7, rotate=0,baseline=-0.1cm,decoration={markings, mark=at position 0.7  with {\arrow{>}}}]
\coordinate (A1) at (90:2);  \coordinate (A2) at (210:2);  \coordinate (A3) at (-30:2);
\coordinate (B1) at (90:1);  \coordinate (B2) at (240:1);  \coordinate (B3) at (-60:1);
\coordinate (C1) at (-150:0.5);  \coordinate (C2) at (-150:1.5); \coordinate (C0) at (-150:1); 
\draw[dashed ] (0,0) circle (2);
\draw  (A1)--(0,0); 
\draw [postaction={decorate}] (0,0)--(A3); 
\draw  [postaction={decorate}]   (A2)--(C2);
\draw  [postaction={decorate}]   (C1)--(0,0  ) ;
\draw[fill=white]  (C0) circle (0.5);
\draw[fill=black!10] (C0) circle(0.5);
\draw[fill=black] (A1) circle (0.07); 
\draw[fill=black] (A2) circle (0.07); \draw[fill=black] (A3) circle (0.07); 
\node[left] at (A2) {$\sft_1$};
\node[right] at (A3) {$\sft_2$};
\node[above] at (A1) {$\sft_3$};
 \end{tikzpicture}\la{4.23} 
 \ee
 may be represented as 
 \be
 \la{4.24}
   \langle\sfPsi(\sft_1) \widetilde{\sfPsi} (\sft_2)\sfPhi(\sft_3) \rangle^{\text{self-energy}}_{\text{1-loop}}
   = \langle\sfPsi(\sft_1) \widetilde{\sfPsi} (\sft_2)\sfPhi(\sft_3) \rangle _{\text{tree}}\times
  \Big(  \frac{   \langle\sfPhi \sfPhi \rangle_{\text{1-loop}} } {   \langle\sfPhi \sfPhi \rangle_{\text{tree}} } 
+2 \frac{   \langle\sfPsi \widetilde\sfPsi \rangle_{\text{1-loop}} } {   \langle\sfPsi \widetilde\sfPsi \rangle_{\text{tree}} } 
  \Big) \ , 
 \ee
 with 
 the  full  1-loop correction to   three-point function   thus given by 
 \be\la{4.25}
    \langle\sfPsi(\sft_1) \widetilde{\sfPsi} (\sft_2)\sfPhi(\sft_3) \rangle _{\text{1-loop}}
    = 
      \langle\sfPsi(\sft_1) \widetilde{\sfPsi} (\sft_2)\sfPhi(\sft_3) \rangle^{\text{triangle}}_{\text{1-loop}}
    +   \langle\sfPsi(\sft_1) \widetilde{\sfPsi} (\sft_2)\sfPhi(\sft_3) \rangle^{\text{self-energy}}_{\text{1-loop}}~.
 \ee
In view of    \rf{4.22}   to \rf{4.10},\rf{4.15}   we    conclude  that  for  the special scheme choice \rf{413}   under which the two-point functions 
 do not receive    one-loop corrections   the same  is  true  also for the   three-point function \rf{4.25}.
 Then comparing to the correlators of   chiral currents in  \rf{2.36},\rf{2.37}  this   suggests that the coefficient $\kappa$  in \rf{2.39} 
 does not  receive quantum 
 corrections.\foot{\ We are assuming   that the  quantum theory  is defined by the  path integral  with the WZW action \rf{2.1},\rf{2.2},\rf{2.5}
 where the overall coefficient $k$  or $b$ in \rf{2.5}  has its classical   value
 (an action with a shifted  $k$   would  correspond to a different scheme choice). 
 It is not clear if  the   quantum effective action \ci{Tseytlin:1993my,deWit:1993qv} 
given by the WZW  action with $k \to k + c_G$ (that reproduces  correlators of currents computed in perturbation theory on a plane)
is a  possible starting point in computing  boundary correlators of elementary fields  of  the WZW theory in  AdS$_2$. 
 }

In Appendix \ref{app:redef}  we will   further 
elaborate on the  issue of the scheme   dependence  of the one-loop corrections to the boundary correlators 
starting with a classically  equivalent  action in terms of redefined   fields. 

\iffa 

\subsection{Comparison   with  correlators of chiral currents} 

For the sake of brevity, let us denote $g(w,w)\equiv g_{0}$.  We have found previously, cf. (\ref{4.10}), that 
the one-loop result for the $\sfPsi$ field 2-point function is 
\be
\la{4.36}
\bt_{12}^{2}\langle \sfPsi(\bt_{1})\widetilde{\sfPsi}(\bt_{2})\rangle = 4\,\Big(1+\frac{1-g_{0}}{2}\,b^{2}+\cdots\Big).
\ee
All-order universality of  $\kappa$ factors, cf. (\ref{2.39}) -- or equivalently the validity at quantum level of the 
symmetry relation (\ref{B.7}) --  implies that also 
\be
\la{4.37}
\bt_{12}^{2}\langle \sfPhi(\bt_{1})\sfPhi(\bt_{2})\rangle = 2\,\Big(1+\frac{1-g_{0}}{2}\,b^{2}+\cdots\Big).
\ee
From the boundary field/WZW current relation 
$\sfPhi=\kappa\,H$ and  (\ref{2.36}) this means
\be
\la{cond1}
k\,\kappa^{2} = 2\,\Big(1+\frac{1-g_{0}}{2}\,b^{2}+\cdots\Big).
\ee
Similarly, using the one-loop correction to the two-point function in \eqref{4.36} and \eqref{4.37}
to compute   the self-energy correction in  \eqref{4.24}, we obtain the  
full loop-corrected three-point function 
\be
\la{4.41}
\bt_{12}\bt_{13}\bt_{23}\langle\sfPhi\sfPsi\widetilde{\sfPsi}\rangle =
-4\,i\,b-8\,i\,(1-g_{0})\,b^{3}+\cdots.
\ee
Comparing with (\ref{2.38}), we get a second matching condition
\be
\la{cond2}
-2\,i\,\sqrt{2}\,k\,\kappa^{3} = -4\,i\,b-8\,i\,(1-g_{0})\,b^{3}+\cdots.
\ee
The solution of (\ref{cond1}) and (\ref{cond2}) reads
\be
\kappa(b) = \frac{b}{\sqrt 2}\,\bigg[1+\frac{3}{2}(1-g_{0})\,b^{2}+\mc O(b^{4})\bigg],\qquad k(b) = 
\frac{4}{b^{2}}\,\bigg[1-\frac{5}{2}(1-g_{0})\,b^{2}+\mc O(b^{4})\bigg].
\ee
This may also be written as 
\be
\kappa = \sqrt\frac{2}{k-2\,(1-g_{0})}+\mc O(b^{5}),
\ee
where we remark that higher order corrections to the perturbative two- and three-point function only
affect the $\mc O(b^{5})$ term. As we have remarked, the choice $g_{0}=1$ is required by the symmetry
condition (\ref{B.7}) -- assuming that relation is not renormalized. Then, $\kappa$ does not receive corrections
and stay at the classical value. Similarly, the tree relation $k=\frac{4}{b^{2}}$ is still valid at this order.

\fi

\section{Boundary correlators  and scattering  amplitudes on  \adst}  \la{sec5} 

While the  scattering amplitudes for the massless WZW
fields in flat space is known to  vanish \cite{Figueirido:1988ct,Hoare:2018jim}, 
we have seen that  the coordinate-space boundary correlators for WZW  fields  \adst are non zero.
Their structure, however,  is simple being dictated by the  KM symmetry. 
One may wonder  if with some natural definition of the AdS  S-matrix  they  may   actually correspond to 
 trivial scattering in \adst or on half-plane. 
 Below we will attempt to clarify this issue.


It is useful first to   recall   what happened  in the   Liouville theory --   how triviality of scattering  
in  \adst emerges in that case.  
The flat space  scattering in this theory was  argued  to be trivial in 
\cite{Thorn:1983qr}, based on previous  results about the energy-momentum eigenstates in finite volume
\cite{Curtright:1982gt,Braaten:1982fr,Braaten:1982yn,Gervais:1982yf}.\foot{\ To avoid infrared problems, the theory may
considered on a circle, where the Liouville field $\vp$ can be expressed in terms
of a free field $\vp^{(0)}$ by means of a quantum B\"acklund transformation. All  
energy-momentum eigenstates on the circle can be obtained by acting on the vacuum
 with the modes of the stress tensor $T^{(0)}_{mn}$ of the B\"acklund field. 
 In \cite{Thorn:1983qr}, it was  argued that the dynamical
 properties of the infinite volume multi-particle states are equivalent to the large radius limit of the (free) 
 $T^{(0)}_{mn}$ eigenstates. This implies that the S-matrix is trivial.} 
The  scattering  in a non-trivial Liouville vacuum   or effectively 
in  \adst   space 
was discussed in \cite{DHoker:1983msr}.\foot{\   
As a normalizable translation-invariant ground state  does not exist in Liouville theory in flat space, 
ref.\cite{DHoker:1983msr}   considered, following \ci{DHoker:1983zwg},     the  theory  in a non-invariant  domain-wall  background 
that spontaneously breaks translation invariance and ``semi-compactifies'' 
 space to a  half-line. The resulting model 
can be identified with the Liouville theory in  \adst geometry.} 
Ref. \cite{DHoker:1983msr}   have shown  that at the tree level there exists a perturbative expansion 
which is infrared safe and  leads to  trivial S-matrix. This conclusion  was generalized   and proved in more formal  way in 
 \cite{Yoneya:1984dd}.

One may  attempt to  define   S-matrix  in AdS  space 
by specifying  suitable ``in'' and ``out'' states  and computing  
amputated bulk correlators (as in flat space LSZ formula). 
In addition to the question of which asymptotic states  to use 
(cf.   \cite{Giddings:1999qu})
a  major technical problem is 
how to explicitly construct the Lorentzian AdS   scattering amplitudes  starting directly from the
 Euclidean  coordinate-space boundary correlators.



 Below we shall  first   outline  the general relation between the AdS 
 scattering amplitudes and the Lorentzian boundary correlators. 
 Then  we shall  discuss the Euclidean $\to$ Lorentzian correlator 
 reconstruction problem in the case of the Liouville theory relating it to the approach of \cite{DHoker:1983msr}.  
 Finally, we shall comment on the  simplest scattering
 amplitude in the WZW theory  in \adst   using  an analogous method. 

\subsection{Massive scalar S-matrix  on  \adst}

 Let us   start   a scalar field  theory in \adst
with  
mass parameter $m^{2}=\Delta(\Delta-1)$. Let us consider 
a Witten  diagram  with one propagator connected to a bulk point $(t, \bz)$
 (here $t$ is real Minkowski time, and $\bz\ge 0$ is the radial 
 \adst    Poincare coordinate).
 Ignoring dependence on   other external points, it may be  symbolically represented 
 as\foot{\ Here $t$  is Minkowski time
 related to    Euclidean \adst time $\sft$   used above by   $t=i \sft $.} 
\be
\la{5.1}
\mc G(t, \bz) = 
\begin{tikzpicture}[line width=1 pt, scale=0.8, baseline=-0.1cm,decoration={markings, mark=at position 0.53 with {\arrow{>}}}]
\coordinate (A1) at (-1.5,0);     \coordinate (A2) at (1,0); 
\draw (A1)--(0,0);
\draw[fill] (A1) circle(0.05); \draw[fill=black!10] (0.8,0) circle(0.8);
\draw[fill] (0,0) circle(0.05); 
\node[left] at (A1) {$(t, \bz)$};
\node[left,yshift=0.4cm,xshift=0.2cm] at (0,0) {$(t', \bz')$};
\node at (0.8,0) {$\Gamma$};
\node[below] at (-0.75,0) {$\mc D$};
\begin{scope}[xshift=0.8cm,yshift=0]
\draw(20:0.8)--(20:1.8);
\draw(-20:0.8)--(-20:1.8);
\draw[thick, dotted] (-18:1.3) arc(-18:18:1.3);
\end{scope}
\end{tikzpicture}
= 
\int dt'\,d\bz' \ \mc D(t, \bz; t', \bz') \Gamma(t', \bz')~,
\ee
where $\mc D \equiv G_{\Delta}$ is the Lorentzian
massive scalar propagator with Dirichlet boundary conditions~\footnote{\ As in (\ref{2.7}), this is for 
the standard normalization of the action, i.e. $S = \frac{1}{2}\int d^{2}x\,\sqrt{g}\,   [(\p\phi)^{2}+m^{2}\,\phi^{2}+\cdots]$.} 
\be
\mc D(t, \bz; t', \bz')  = \tfrac{\mc C_{\Delta}}{(2u)^{\Delta}}\, {}_{2}F_{1}(\Delta, \Delta, 2\Delta, -\tfrac{2}{u})~,
\qquad \mc C_{\Delta} = \tfrac{\Gamma(\Delta)}{2\sqrt\pi\,\Gamma(\Delta+1/2)}~,
\quad u(x,x') = \tfrac{(\bz-\bz')^{2}-(t-t')^{2}}{2\,\bz\,\bz'}~,
\ee
and $\Gamma$  stands for the rest of  the  diagram (i.e. with one line amputated). 
The propagator $\mc D$ may be written as 
\be
\la{5.3}
\mc D(t, \bz; t', \bz') = \frac{1}{2}\int_{0}^{\infty}\frac{d\omega}{\omega}e^{-i\omega|t-t'|}f_{\omega}(\bz)f_{\omega}(\bz')~,
\ee
where the functions $\{f_{\omega}(\bz)\}_{\omega>0}$ are  eigenmodes of the  kinetic  operator for a scalar field in \adst
\be\la{5.4}
\Big(\p_{\bz}^{2}+\omega^{2}-\frac{m^{2}}{\bz^{2}}\Big)\,f_{\omega}(\bz) = 0~.
\ee
They form  a basis in $\bz\in [0,\infty)$ with normalization
\be
\la{5.5}
\int_{0}^{\infty} d\omega\, f_{\omega}(\bz)f_{\omega}(\bz') = \delta(\bz-\bz')~,\qquad\qquad 
\int_{0}^{\infty}d\bz\, f_{\omega}(\bz)f_{\omega'}(\bz) = \delta(\omega-\omega') ~. 
\ee
$f_{\omega}(\bz)$  can be identified with the wave function of the asymptotic 
state with energy $\omega$ created by the scalar field. Its  explicit form  for the Dirichlet boundary condition 
is 
\be
\la{5.6}
f_{\omega}(\bz) = a(\omega)\,\sqrt{\bz}\,J_{\Delta-\frac{1}{2}}(\omega\,\bz)~,
\ee
where the normalization $a(\omega)$ is determined by (\ref{5.5}).  
The corresponding  scattering amplitude $\mc A(\omega_{1}, \dots, \omega_{N})$  may be 
formally defined as 
\begin{align}
\la{5.7}
\mc A(\omega_{1}, \dots, \omega_{N}) =
\int \Big(\prod_{i=1}^{N} dt_{i}\,d\bz_{i}\,e^{i\,\omega_{i}\,t_{i}}\,f_{\omega_{i}}^{(\Delta_{i})}(\bz_{i})\Big)\,
\Gamma(t_{1}, \bz_{1}; \dots; t_{N}, \bz_{N})~,
\end{align}
where in $\Gamma$  we included the  external leg labels  and  the subscript  in 
$f_{\omega}^{(\Delta)}(\bz)$ is (\ref{5.6}) indicates the  corresponding  value of $\Delta$
 (in the case of multi-scalar  scattering with different masses).

 \subsubsection{Comments on relation to boundary correlators}
 
It is possible to  formally ``derive''  a relation between (\ref{5.7}) and a Fourier transform of the   coordinate-space 
 boundary correlators.  Let us consider one leg in \rf{5.1}   taken to the boundary, i.e. define 
the boundary correlator  
\be
\la{5.8}
\rA(t) = \lim_{\bz\to 0}\bz^{-\Delta}\mc G(t, \bz) = 
\begin{tikzpicture}[line width=1 pt, scale=0.8, baseline=-0.1cm,decoration={markings, mark=at position 0.53 with {\arrow{>}}}]
\coordinate (A1) at (-1.5,0);     \coordinate (A2) at (1,0); 
\draw (A1)--(0,0);
\draw[thin, densely dashed] (A1) arc(180:140:1.5);  \draw[thin, densely dashed] (A1) arc(180:220:1.5);
\draw[thin] (A1) arc(180:160:1.5);  \draw[thin] (A1) arc(180:200:1.5);
\draw[fill] (A1) circle(0.05); \draw[fill=black!10] (0.8,0) circle(0.8);
\draw[fill] (0,0) circle(0.05); 
\node[left] at (A1) {$t$};
\node[left,yshift=0.4cm,xshift=0.2cm] at (0,0) {$(t', \bz')$};
\node at (0.8,0) {$\Gamma$};
\node[below] at (-0.75,0) {$\mc D$};
\begin{scope}[xshift=0.8cm,yshift=0]
\draw(20:0.8)--(20:1.8);
\draw(-20:0.8)--(-20:1.8);
\draw[thick, dotted] (-18:1.3) arc(-18:18:1.3);
\end{scope}
\end{tikzpicture}~,
\ee
where the  circular line on the left  denotes \adst boundary.
Substituting (\ref{5.3}) into (\ref{5.1}) and 
computing (\ref{5.8}) using that  $f_{\omega}(\omega\bz) \sim \bz^{\Delta}$ for $\bz\to 0$, we find 
that the Fourier transform of $\rA(t)$ is actually the same as the scattering amplitude in (\ref{5.7}). Indeed,  
(here $c_{\Delta}$ is a coefficient dependent only on $\Delta$) \footnote{\ Here and in the following, integrals $\bz$ are restricted to the 
 \adst  region, i.e.  $\bz\ge 0$.}
\begin{align}
 \mc A(\omega) \equiv \int dt\,e^{i\omega\,t}\, \rA(t) \notag &= c_{\Delta}\,\int dt e^{i\omega t}\int dt'\,d\bz'\int_{0}^{\infty}
\frac{d\omega'}{\omega'}e^{-i\omega'|t-t'|}a(\omega')
\omega'^{\Delta}
\,f_{\omega'}(\bz')\,\Gamma(t', \bz') \notag \\ 
&= c_{\Delta}\int dt'\,d\bz' e^{i\omega t'}\,\int_{0}^{\infty}\frac{d\omega'}{\omega'}\frac{i\,a(\omega')\,\omega'^{\Delta}}{\omega^{2}-\omega'^{2}}
\,f_{\omega'}(\bz')\,\Gamma(t', \bz')~.\la{59}
\end{align}
Evaluating the integral over  $\omega'$ by 
picking (one-half of) the contribution from the pole at $\omega'=\omega$, we obtain
\be
\la{5.10}
\mc A(\omega) = \mc N(\omega)\,\int dt\,d\bz \ e^{i\omega t} f_{\omega}(\bz)\,\Gamma(t, \bz)~,
\ee
in agreement with (\ref{5.7}).
The same result is found by directly considering the boundary limit of (\ref{5.1}). This amounts to 
replacing the bulk propagator $\mc D$ by the bulk-to-boundary expression
\be
\la{5.11}
\rA(t) = \lim_{\bz\to 0}\bz^{-\Delta}\mc G(t, \bz) = \mc C_{\Delta}\,
\int dt'\,d\bz' \ \left[\frac{\bz'}{-(t-t')^{2}+\bz'^{2}}\right]^{\Delta} \Gamma(t', \bz')~,
\ee
and taking the  Fourier transform of (\ref{5.11}) in Cauchy principal value sense (i.e. summing  half  of the two 
residues at $t=t'\pm \bz'$).~\footnote{\ At this  stage   
 this is just a formal prescription. More precisely, one should 
shift the integration contour by adding causal $i\eps$ shifts, see  below.} 
For  $\omega>0$, it reads
\be
\mc C_{\Delta}\fint dt \, e^{i\omega t} \left[\frac{\bz'}{-(t-t')^{2}+\bz'^{2}}\right]^{\Delta} = 
\frac{2^{-\frac{1}{2}-\Delta}\,\pi}{\Gamma(\Delta+1/2)}\,\omega^{\Delta+\frac{1}{2}}\,\sqrt{\bz}\,
J_{\Delta-1/2}(\omega\,\bz)\,
e^{i\omega t'}~.
\ee
and thus implies again (\ref{5.10}). 

To summarize, we have shown that under 
a certain prescription, one can start with  the   $N$-leg boundary correlator for fields with dual conformal dimensions 
$\Delta_{1}, \dots, \Delta_{N}$
\be
\rA(t_{1}, \dots, t_{N}) = \lim_{\bz_{i}\to 0} \bz_{1}^{-\Delta_{1}}\cdots
\bz_{N}^{-\Delta_{N}}\, \int \Big(\prod_{i=1}^{N}dt'_{i}d\bz'_{i}\
 \mc D(t_{i}, \bz_{i}; t'_{i}, \bz'_{i})\Big)\,\,\Gamma(t'_{1}, \bz'_{1}; \cdots; t'_{N}, \bz'_{N})~,
\ee
take its Fourier transform in each leg and as result   find  an alternative  representation  for the 
scattering amplitude $\mc A(\omega_{1}, \dots, \omega_{N})$ in (\ref{5.7}), i.e.
\begin{align}
\la{5.14}
\mc A(\omega_{1}, \dots, \omega_{N}) &= \int \Big(\prod_{i=1}^{N} dt_{i}e^{i\,\omega_{i}\,t_{i}}\Big)\,
\rA(t_{1}, \dots, t_{N})~.
\end{align}

Let us note that 
 the amputated Green's function $\Gamma$ in (\ref{5.7}), as well as the boundary correlator in (\ref{5.14}),
are  the Lorentzian ones. In general, the explicit analytical continuation of the boundary correlators 
from the Euclidean to the Lorentz signature 
should be done according to the general prescriptions based on reconstruction theorems  \cite{Luscher:1974ez}
as discussed more recently in \cite{Hartman:2015lfa,Bautista:2019qxj,Gillioz:2019lgs}. 
In particular, to compute the fully time-ordered Wightman function from the Euclidean correlators, 
one replaces $t_{i}\to t_{i}-i\eps_{i}$ with $\eps_{i}>\eps_{j}$ when  $t_{i}>t_{j}$ and then takes $\eps_{i}\to 0$.~\footnote{\ 
As first discussed in \cite{Mack:2009mi},
the analytical continuation can be done at the level of Mellin amplitudes, see  \cite{Mack:2009gy,Penedones:2010ue,Fitzpatrick:2011ia}. 
}
The Fourier transform of the resulting expression 
is expected to give the scattering amplitude and to match (\ref{5.7}). 

\subsubsection{Tree level scattering in Liouville theory on  \adst}

To illustrate the  relation between   (\ref{5.7}) and (\ref{5.14})  let us    consider again the 
Liouville theory following \cite{DHoker:1983msr} . The basic $1\to 2$ particle production process  $\varphi\to \varphi+\varphi$ 
here  is particularly simple:  at tree level 
it involves the amputated 
3-point function that is just a constant. Let us begin by (\ref{5.7}). The off-shell wave functions  (\ref{5.6})  are 
\be\la{515}
f_{\alpha, \omega}(\sft,\sfz)=e^{ i\alpha t}    \sqrt{\omega\sfz} J_ {3/2} (\omega \sfz)~, \qquad \omega>0~.
\ee
The on-shell condition is $\alpha^2=\omega^2$, namely $\alpha=\pm\omega$.
 Besides, $f_{-\omega, \omega}(\sft,\sfz)= f_{-\omega, -\omega}(\sft,\sfz)$,
and we can simultaneously treat both signs of $\omega$, i.e. ``in'' or ``out'' states. Up to irrelevant constants, 
the scattering amplitude for a 3-particle process may be written as 
\be
\cA_3(\alpha_1,\omega_1;\alpha_2,\omega_2;\alpha_3,\omega_3    )\sim \delta(\alpha_{1}+\alpha_{2}+\alpha_{3})\, \bar \cA_3 (\omega_1,\omega_2,\omega_3)~,
\ee
where
\be
\la{5.17}
\bar \cA_3 (\omega_1,\omega_2,\omega_3)
=\int_0^\infty \frac{d\sfz}{\sfz^2} f_{\omega_1} (z)f_{\omega_2} (z)f_{\omega_3 } (z)
=\sqrt{\omega_1 \omega_2 \omega_3 }  \int_0^\infty \frac{d\sfz}{\sqrt\sfz }
J_{3/2} (\omega_1 \sfz)J_{3/2} (\omega_2 \sfz)J_{3/2} (\omega_3 \sfz) ~. 
\ee
We  may now use  the  known value of the  following definite integral\footnote{\ Useful integrals involving
 three Bessel functions are discussed 
in \cite{Gervois:1985ff,Gervois:1985fe}.}
\be
\la{5.18}
  \int_0^\infty \frac{d\sfz}{\ \sfz^{\nu-1} }
J_{\nu} (\omega_1 \sfz)J_{\nu} (\omega_2 \sfz)J_{\nu} (\omega_3 \sfz)
=\frac{2^{\nu-1} {\rS}^{2\nu-1}}{\sqrt\pi\,(\omega_1\omega_2\omega_3)^\nu \Gamma(\nu+\frac12)}~,
\ee
where 
\be
\rS=\frac14 \sqrt{(\omega_1+\omega_2+ \omega_3) (-\omega_1+\omega_2+ \omega_3) 
(\omega_1-\omega_2+ \omega_3) (\omega_1+\omega_2- \omega_3)  }~,
\ee
is the area of a triangle with sides $\omega_1,\omega_2,\omega_3$  (if $\omega_{1}, \omega_{2},\omega_{3}$
do not form a triangle, the integral is zero).  From (\ref{5.18})  ref.\cite{DHoker:1983msr}  found the following expression for \rf{5.17} 
\be
\la{5.20}
\bar \cA_3 =\sqrt{\frac{2}{\pi}} \frac{\rS^2}{\omega_1 \omega_2 \omega_3}~.
\ee
As the kinematically allowed 3-particle processes are associated with a degenerate triangle with vanishing area $\rS=0$
one finds that  $\cA_{3}=0$. This 
calculation has been extended in \cite{DHoker:1983msr}  to the 
4-particle  scattering processes  that were also  found to vanish.

To try to  recover (\ref{5.20}) as a  Fourier transform  (\ref{5.14}) of the boundary correlator 
we need first  to analytically continue the Euclidean boundary 3-point function 
$\sim \frac{1}{\sft_{12}^{2}\sft_{13}^{2}\sft_{23}^{2}}$ to the  Lorenzian signature 
(to get the Lorentzian  time-ordered 3-point function). 
Evaluating the associated Fourier transform seems far from trivial
because the $d^{3}t$ integration region has to be split according to the time ordering and suitable $\pm i\eps$ shifts have to be 
introduced.\footnote{\ The Fourier representation of the 
Wightman Lorentzian 3-point function with fixed time ordering $\langle \mc O(t_{1})\mc O(t_{2})\mc O(t_{3})\rangle$, 
$t_{1}>t_{2}>t_{3}$ is  discussed  in \cite{Bautista:2019qxj}.}
In principle, 
another approach is to look for an analytic continuation 
of the triple-$K$ integral representation of the Euclidean 3-point function \cite{Bzowski:2013sza}. 
Such analytic continuations have been recently discussed in \cite{Lipstein:2019mpu}. 

Let us note that  continuation to Lorentzian 
signature and time-like momenta requires an analytic continuation of expressions involving the Appel function and this 
 is known to be related  to triple-$J$ integrals for special arguments, see Eq.~(7.1) of  \cite{bailey1936some}
 and also \cite{Gervois:1985ff,Gervois:1985fe}. This procedure is yet to be investigated in detail, but  let us note 
 that  
 \be
 \la{521}
  \int_0^\infty d\sfz\,\sfz^{1-\nu+\eps}\,
K_{\nu} (\omega_1 \sfz)\,K_{\nu} (\omega_2 \sfz)\,K_{\nu} (\omega_3 \sfz)
=\frac{\pi^{2}}{2\,\eps}\,
\frac{2^{\nu-1} {\rS}^{2\nu-1}}{\sqrt\pi\,(\omega_1\omega_2\omega_3)^\nu \Gamma(\nu+\frac12)}+\mc O(1)~.
\ee
This relation shows that with a simple (although ad hoc) regularization of the triple-$K$ integral, the 
triple-$J$ integral (relevant for the  scattering amplitude)  shows up 
as the residue at the singular pole. 
The fact that leading singularities of divergent triple $K$ integrals may contain physical objects
has been discussed in the  Euclidean context in \cite{Bzowski:2015pba}. It would be interesting to understand the
relation between their analysis and relations like (\ref{521}).


 \subsection{Massless  scattering case}
 
 In view  of  the subtleties involved  in extracting  the scattering amplitudes from the Euclidean boundary correlators, 
 here we shall  consider  massless scattering  following the approach of   \cite{DHoker:1983msr}  based on (\ref{5.7}). 
 Let us start with the   simplest 
  $1\to 2$ process  and   emphasize the difference  between  models   with derivative-independent  scalar $\phi^3$   vertex and 
 with  $\phi (\p \phi)^2$   \sm type   (classically) conformally invariant vertices.

 For a massless scalar we have  $\Delta=1$ or  $\Delta-\ha=\frac{1}{2}$  in \rf{5.6} and  for a   $\phi^3$   interaction vertex 
 the analog of the integral in \rf{5.17}  representing  the tree  level  $1\to 2$  particle production  amplitude is  
  \begin{align}
 \la{5.22}
&  \int_0^\infty \frac{d\sfz}{\ \sqrt{\sfz} }
J_{1/2} (\omega_1 \sfz)J_{1/2} (\omega_2 \sfz)J_{1/2} (\omega_3 \sfz) \notag \\
&= \frac{1}{2\sqrt 2\,\pi^{3/2}\,\sqrt{\omega_{1}\omega_{2}\omega_{3}}}\,\Big[
\Omega\,\log(\Omega^{2})-\omega_{12}\,\log(\omega_{12}^{2})-\omega_{13}\,\log(\omega_{13}^{2})-\omega_{23}\,\log(\omega_{23}^{2})
\Big]~,
\end{align}
where we defined $\omega_{12} = \omega_{1}+\omega_{2}-\omega_{3}$,  {\em etc.},   and $\Omega=\omega_{1}+\omega_{2}+\omega_{3}$.
The integral (\ref{5.22}) does not vanish on-shell. For instance, if $\omega_{3}\to \omega_{1}+\omega_{2}$ it has a finite non-zero limit.
 \footnote{\ 
Notice that we can put an arbitrary scale $\mu$ in the logarithms in (\ref{5.22}) since $\Omega-\omega_{12}-\omega_{13}-\omega_{23}=0$.
}

In the \sm case  
in flat space the 3-point amplitude   vanishes due to on-shell kinematics. 
This is less  automatic   in the \adst case. 
Let us consider  the case of a general \sm  in the parametrization used in \rf{3.2},\rf{3.8}  where  the cubic  vertex in the 
WZW  case is $\sim  f_{abc}\p X^{a}\pb X^{b}\, X^{c}$.
Because of antisymmetry  of $f_{abc}$  the  vertex  is effectively  $\sim \ha (  \p X^{a}\pb X^{b} -    \pb X^{a}\p X^{b}) X^c $.
Let us first consider the contribution of the first term   and then antisymmetrize in momenta. 
We will   need  the wave functions
\begin{align}
\la{5.23}
f_{\pm\omega, \omega}(t,\sfz) &=  e^{\pm i \omega t }\sin (\sfz \omega)~,  \\
\p  f_{\pm\omega, \omega}(t,\sfz) &\propto(\p_t+\p_\sfz)e^{\pm i \omega t }\sin (\sfz \omega) 
=\omega e^{\pm i \omega(t+\sfz)}~, \no \\
\bar\p  f_{\pm\omega,\omega}(t,\sfz)  &\propto(\p_t-\p_\sfz)e^{\pm i \omega t }\sin (\sfz \omega) 
=- \omega e^{\pm i \omega(t-\sfz)}~.\no 
\end{align}
Starting   from   \rf{5.7}, suppressing the group indices     and defining   $\alpha_{i}=\pm \omega_{i}$
  we find  (cf. \rf{5.17},\rf{5.22})
\begin{align}
\cA'_{3} &= \int  d t d\sfz\; \p f_{\alpha_{1},\omega_1}  (t,\sfz) \bar\p f_{\alpha_{2},\omega_2} (t,\sfz)
f_{\alpha_{3},\omega_3 } (t,\sfz)
\propto  \int d t d\sfz\;  \omega_1 \omega_2 
e^{i \alpha_{1}(t+\sfz)}e^{i \alpha_{2}(t-\sfz)} e^{i \alpha_{3} t }\sin (\sfz \omega_3) \notag \\
&\propto   \delta(\alpha_{1}+\alpha_2+\alpha_3)\, \omega_1 \omega_2  \,  \int_0^\infty  d\sfz 
\Big[ e^{i (\alpha_{1}-\alpha_{2}+\alpha_{3})\sfz}  -e^{i (\alpha_1-\alpha_{2}-\alpha_3)\sfz}   \Big] \notag \\
&= \delta(\alpha_1+\alpha_2+\alpha_3)\, \omega_1 \omega_2   
\Big[  \pi\delta (\alpha_1-\alpha_2+\alpha_3)   +\frac{i}{ \alpha_1-\alpha_2+\alpha_3}
- \pi\delta (\alpha_1-\alpha_2 -\alpha_3) -\frac{i}{\alpha_1-\alpha_2-\alpha_3}
\Big]\notag \\ 
& \propto \delta(\alpha_1+\alpha_2+\alpha_3)\,\omega_{1}\,\omega_{2} \,\frac{\alpha_{1}+\alpha_{2}}{\alpha_{1}\,\alpha_{2}}~.\la{523}
\end{align}
Here we used that 
$\int_0^\infty d\sfz\,  e^{i \omega \sfz}  =\pi \delta(\omega) +{i}{\omega}^{-1}$ .
We are still  to antisymmetrize in $\omega_1\leftrightarrow \omega_2$,  but since  the expression in \rf{523}  is symmetric, the final result is thus zero. 
Thus the  3-point scattering amplitude  vanishes also in AdS$_2$.\footnote{\ Let us note that 
dealing with massless 2d fields  requires   extra care.    
The wave function $f$ in (\ref{5.23})
is not vanishing for $\bz\to \infty$. Thus, integration by parts is not  a priori allowed in (\ref{523}).
That means  that  the starting  form of the action   may be important as  the contribution of boundary terms 
(produced by integrations by parts)  may  be non-trivial.}

\iffa 
Let us consider the 3-point scattering amplitude on the example 
of  the   $SL(2, \mathbb R)$  WZW  model  discussed in section  2. 
 To  compute  the amplitude $\psi+\tilde \psi \to \phi$ in the $SL(2, \mathbb R)$ WZW model we  need 
  to use  the 2-derivative vertex in \rf{2.5}  and 
\begin{align}
\la{5.24}
\cA_{3} &= \int  d t d\sfz\; \p f_{\omega_{1},\omega_1}  (t,\sfz) \bar\p f_{\omega_{2},\omega_2} (t,\sfz)
f_{-\omega_{3},\omega_3 } (t,\sfz)
\propto  \int d t d\sfz\;  \omega_1 \omega_2 
e^{i \omega_1(t+\sfz)}e^{i \omega_2(t-\sfz)} e^{-i \omega_3 t }\sin (\sfz \omega_3) \notag \\
&\propto   \delta(\omega_1+\omega_2-\omega_3)\, \omega_1 \omega_2  \,  \int_0^\infty  d\sfz 
\Big[ e^{i (\omega_1-\omega_2+\omega_3)\sfz}  -e^{i (\omega_1-\omega_2-\omega_3)\sfz}   \Big] \notag \\
&= \delta(\omega_1+\omega_2-\omega_3)\, \omega_1 \omega_2   
\Big[  \pi\delta (\omega_1-\omega_2+\omega_3)   +\frac{i}{ \omega_1-\omega_2+\omega_3}
- \pi\delta (\omega_1-\omega_2 -\omega_3) -\frac{i}{\omega_1-\omega_2-\omega_3}
\Big]\notag \\ 
& \propto \delta(\omega_1+\omega_2-\omega_3) 
( \omega_1+\omega_2 ) \propto \delta(\omega_1+\omega_2-\omega_3) \,\omega_{3},
\end{align}
where we used that 
\be
\int_0^\infty e^{i \omega \sfz} d\sfz =\pi \delta(\omega) +\frac{i}{\omega}\ .
\ee
So, similarly to the case of a cubic interaction, even in this conformally invariant case, we have a non-zero 3-particle
on-shell amplitude. Notice however that 
here the parametrization
is a bit special since the 3-point vertex comes  both from the metric and $B_{\mu\nu}$ term.

\fi



As for the 
 4-particle   scattering amplitude, 
 in the Liouville theory in \adst it   was 
  found to 
     vanish in a non-trivial way,   due to a    cancellation of  different contributions \cite{DHoker:1983msr}.
It would   be interesting to see  if  it   also   vanishes      in the WZW  theory in AdS$_2$. 
A possible   reason  of  why this   may happen   is 
the absence of  non-trivial  structures in the corresponding  Euclidean boundary correlators, i.e. the cancellation of logarithmic
terms that  happens in the Liouville theory \ci{Ouyang:2019xdd,Beccaria:2019stp} 
and that we also
observed  above   for the  WZW   limit of a general $\s$-model.  
To establish this   link  it remains to 
 derive the \adst scattering amplitudes
from Euclidean boundary correlators in a   systematic way. 

\section{Concluding remarks} \la{sec6}

In this paper we considered  boundary correlators  of elementary  fields of 2d  $\sigma$-models in AdS$_2$. 
Similar  problem  
appears in the  study of correlators of operators on a Wilson line  in the  strong-coupling   description in terms of the AdS$_5 \times S^5$  Nambu  string action
 in the static gauge \ci{Giombi:2017cqn,Beccaria:2019dws}. 
 One motivation  is to learn how to compute  loop Witten diagrams in \adst  in models   with derivative interactions. 
 We have observed, in particular, 
  that the  structure of  four-point   correlators  simplifies (with logarithmic terms  of the 1d cross-ratio  cancelling out) only 
 in the WZW  case when the \sm   has an extra KM symmetry. 
 In that case the boundary correlation functions  of the WZW  fields  are found to  be the same as  the correlators 
 of the chiral WZW  currents on the plane  restricted to the real line. 
 
 Another possible motivation  is related to the search  for new integrable 2d $\sigma$-models using 
 S-matrix based criteria   as  in  the 
  massive  case.  
If  one expands  near a trivial \sm vacuum in flat 2d space 
 one gets  massless   scattering amplitudes  which, in general,   suffer   from 
 IR ambiguities   \cite{Hoare:2018jim,Donahue:2018bch}.  If instead one considers the   \sm  on \adst then 
 its coordinate-space boundary correlators are better  defined  and one may
 try to  find   the  analogs of  the  standard  integrability  constraints (S-matrix factorization and no particle creation) 
 directly in terms of them.
  As  any  2d   \sm   is classically Weyl invariant, 
 the tree-level problem in \adst  is  equivalent to   the same problem on  flat half-plane  with  particular (Dirichlet) boundary conditions. 
 Hidden  conserved charges   that   exist in a  classically  integrable \sm on a plane should 
 lead to constraints on the  corresponding   Euclidean boundary correlators and  the associated  S-matrix  on  half-plane.
 This should  also extend  to the quantum level if  the \sm is  quantum scale  invariant  (like the WZW model).


\section*{Acknowledgments}

We would like to thank  D. Ponomarev,  S. Giombi  and R. Metsaev   for 
useful  discussions of related questions. 
MB was supported by  the INFN grant GSS (Gauge Theories, Strings and Supergravity).
HJ was supported by Swiss National Science Foundation.
AAT was supported by the STFC grant ST/P000762/1.

\appendix
\section{Notation and conventions}
\la{app:notation}


The AdS$_2$ metric   is 
\be\la{A.1}
ds^2 =  \frac{d\sft^2 +d\sfz^2 }{\sfz^2}=  -4 \frac{ dw d\bar w }{(w-\bar w)^2} ~, \qquad w=\sft+i\sfz~, \qquad \sfz>0~,
\ee
and we use the conventions
  \be\la{A.2} 
  \p \equiv \p_w =\frac12 (\p_\sft -i \p_\sfz)~, \qquad\qquad   \bar\p \equiv \p_{\bar w} =\frac12 (\p_\sft +i \p_\sfz)~,
  \ee
  \be\la{A.3} 
    \epsilon^{\sft \sfz}=-    \epsilon^{\sfz \sft}= \sfz^2~, \qquad 
    \epsilon^{w\bar w}= -   \epsilon^{ \bar w w}=-2i \sfz^2~, \qquad
{\sf g}^{\sfz\sfz}=    {\sf g}^{\sft\sft}=\sfz^2~, \qquad     {\sf g}^{w\bar w}={\sf g}^{\bar w  w}=2\sfz^2~.
  \ee
We also define the integration  measure as follows 
\be\la{A.4} 
d^2 w=d\sfz \, d\sft, \qquad\qquad   {\rm d}^2 w =\frac{d^2 w}{\pi}~.  
\ee
Our  convention for the $\delta$-function is 
\be
\la{A.5}
\delta^{(2)}(w)  =\delta(\sft) \delta(\sfz)~, \qquad
\int d^2 w \,  \delta^{(2)}(w) f(w) =f(0)~, \qquad d^2 w= d\sft d\sfz~, \quad w=\sft+i\sfz~,
\ee
so that  one has 
\be
\la{A.6}
\p \frac{1}{\bar w} =\pi \delta^{(2)}(w) , \qquad \bar\p \frac{1}{ w} =\pi \delta^{(2)}(w) ~.
\ee
The bulk propagator of a  massless field in \adst  with the  action normalized as 
$S=   \int_{\rm AdS_{2}} \rmd^2 w \;  \p\phi \bar\p\phi$ is given by 
\be
\la{A.7}
g(\eta)= -\frac{1}{2} \log \eta(w,w')~  , 
\ee
where the geodesic distance $\eta$ is defined in (\ref{2.8}).
The associated bulk-to-boundary propagator is
\be
\la{A.8}
\gb (\sft ; w' ) = \lim_{\sfz\to 0 } \frac{1}{\sfz} g  (\sft,\sfz ;\sft' ,\sfz' )
= \frac{2\sfz' }{ (\sft'-\sft)^2 + \sfz'{}^2}
 = \frac{-i  }{  \sft-w'  }+ \frac{ i  }{  \sft- \bar w'  }~.
  \ee

\section{Global  symmetry  constraints  in  $SL(2,\mathbb R)$  WZW model}   
\la{app:symm}


Let us consider the consequences of the global invariance of the WZW action 
\rf{2.5}  under $g\to U g$ where $U$ is a $SL(2,\mathbb R)$ matrix that may be chosen as    
\be\la{b1}
U = \begin{pmatrix}
1+\gamma & \rho \\
\eps & 1-\gamma 
\end{pmatrix} \ , 
\ee
where $(\gamma, \rho, \eps)$   are constant parameters. The infinitesimal transformation
of the fields in (\ref{2.2}) reads
\be
\la{b2}
\delta\phi = \sqrt{k}\,(-\gamma+\psi\eps)~,\qquad
\delta\psi = 2\psi\gamma+\rho-\psi^{2}\eps~,\qquad
\delta\tilde\psi = k\,e^{-\frac{2}{\sqrt k}\phi }\,\eps~.
\ee
In particular, taking  $\gamma=\rho=0$ and rescaling $\eps \to b\,\eps$  where $b= {2 \ov \sqrt k}$ gives  
%
%
\iffa \be\la{b2}
\delta\tilde \psi = e^{-2\,\phi}\,\theta_{3},\qquad
\delta\psi = 2\psi\theta_{1}+\theta_{2}-\psi^{2}\theta_{3},\qquad
\delta\varphi = -\theta_{1}+\psi\theta_{3} \ , 
\ee
\fi 
\be
\la{b3}
\delta\phi = 2\psi\,\eps~, \qquad \delta\psi = -b\,\psi^{2}\,\eps~,\qquad \delta\psib = {4}{b}^{-1} \,e^{-b\phi}\,\eps \ . 
\ee
The action \rf{2.5} is readily checked to be  invariant under \rf{b3}  (using integration by  parts). 
 Using the boundary asymptotics \rf{2.6}   we get   from the   $\bz\to 0$ limit of (\ref{b3})
 the following transformation of the corresponding boundary fields  
 \begin{align}\la{b4} 
 \delta\sfPhi(\bt) = 2\,\sfPsi(\bt)\,\eps + \mc O(\eps^{2}, \bz)~,\qquad 
 \delta\sfPsi(\bt) = \mc O(\bz)~,\qquad 
 \delta\widetilde{\sfPsi}(\bt) = 4 \big[ b^{-1} \bz^{-1} - \sfPhi(\bt)\big] \,\eps+\mc O(\eps^{2},\bz)~.
 \end{align}
 Assuming the computational scheme preserves   the  global  $SL(2,\mathbb R)$  symmetry, it then 
  imposes constraints on the 
 boundary   correlators.
    In view of the   symmetry rotating  $\psi$  into  $\psib$  one  should have 
  $\langle\sfPhi(\bt_{1})\widetilde{\sfPsi}(\bt_{2})\rangle=0$. 
Applying the variation \rf{b4} to this relation gives 
 \be\la{b5} 
0=2\langle\sfPsi(\bt_{1})\widetilde{\sfPsi}(\bt_{2})\rangle+4 \langle\sfPhi(\bt_{1})
4 \big[ b^{-1} \bz^{-1} - \sfPhi(\bt_2)\big]  \rangle
+\mc O(\eps, \bz)~.
 \ee 
The $SL(2,\mathbb R)$    symmetry   implies   that the 
 tadpole  $\langle \sfPhi\rangle $   should vanish  ($\phi$  is shifted by  the parameter $\theta_1$  in  \rf{b1}).\footnote{\ \la{f32}
Note that the  one-loop  contribution to $\langle\sfPhi\rangle$  given by  the 
  tadpole  with $(\psi, \tilde \psi)$ propagator  computed  with a cutoff $\bz>\eps$ is  linearly  divergent 
 \be\no 
\begin{tikzpicture}[line width=1 pt, scale=0.8, baseline=-0.1cm,decoration={markings, mark=at position 0.53 with {\arrow{>}}}]
\coordinate (A1) at (-0.8,0);        \coordinate (A2) at (0.8,0); 
\coordinate (B1) at (-2,0);        \coordinate (B2) at (2,0); 
\node[above] at (B1) {$\bt_{1}$};  
\draw (A1)--(B1); 
\draw[postaction=decorate] (A1) arc(-180:180:0.6);
\end{tikzpicture} \
\sim \int d^2w\ \frac{\bz}{(\bt-\bt_{1})^{2}+\bz^{2}}\frac{1}{(2\,i\,\bz)^{2}} 
\to -\frac{1}{4}\int_{-\infty}^{\infty}dt\ \int_{\eps}^{\infty} d\bz \ \frac{1}{\bz\,(\bt^{2}+\bz^{2})} = -\frac{\pi}{4\,\eps}~.
\ee
This divergence is to be subtracted in a $SL(2,\mathbb R)$ preserving scheme (see also discussion below  \rf{4.1}). 
}
We  thus find the following relation 
\be
\la{B.7}
\langle\sfPsi(\bt_{1})\widetilde{\sfPsi}(\bt_{2})\rangle = 2\,\langle\sfPhi(\bt_{1})\,\sfPhi(\bt_{2})\rangle \ .  \ee 
This relation  is expected to hold at  the quantum level  assuming the 
 above 
 $SL(2, \mathbb R)$ symmetry is preserved by the  computational scheme. 
 This is  a necessary condition for matching the    correlation functions of  chiral currents on  which 
 $SL(2, \mathbb R)$   acts linearly. 
 

 \section{Alternative computation of  one-loop   boundary correlators 
 in  $SL(2, \mathbb R)$    WZW model}  \la{appC}
 \la{app:redef}

Here we shall revisit   the computation of the one-loop corrections to the two-point  boundary correlators 
in  $SL(2, \mathbb R)$    WZW model discussed in section~\ref{1-loop2pt}. 
We shall use an alternative  form of the action  in terms of  redefined  field  variables.
Local  field redefinitions are, in general,   expected to leave  the  physical (boundary)  correlators  
invariant
provided  they are properly defined (taking into account  wave-function  renormalization factors, etc).\foot{\ This is easy to see at the tree  level:  
redefinitions like $\varphi \to \varphi + \varphi^2 + ... $ 
with $\varphi$ subject to the boundary conditions like \rf{2.6}   produce terms of higher order in $\bz\to 0$  in the  correlators.} 
Here we shall  first    follow a naive  approach  ignoring  this subtlety. 


Let us start   with the action \rf{2.5}   and represent it     in terms of the  
redefined 
fields $(\chi, \widetilde \sPsi)$ defined by 
\be\la{c1}
\psi =e^{-b \phi /2} \sPsi~, \qquad\qquad  \tilde\psi =e^{-b \phi /2}  \widetilde \sPsi~.
\ee
Then   up to  the quartic terms  \eqref{2.5} is given by \footnote{\ Note that the cubic term can be rewritten as 
$
\sPsi\bar\p \widetilde \sPsi \p \phi +\widetilde\sPsi\p \sPsi \bar\p \phi 
= \sPsi (\bar\p \widetilde\sPsi \p \phi - \p \widetilde\sPsi \bar\p \phi)  -  \sPsi\widetilde\sPsi\bar\p\p \phi .
$
The first two terms here represent  the standard  WZ term, while the last term can be removed  by a 
  redefinition  of $\phi$  under which  an extra quartic term will  be generated. 
}
\beqn
\la{C3}
S&=&
 \int \rmd^2 w \Big[     \p\phi \bar\p\phi   +      \p \sPsi \bar\p \widetilde\sPsi  -\frac{b}{2} (\sPsi\bar\p \widetilde\sPsi \p \phi +\widetilde\sPsi\p \sPsi \bar\p \phi) 
+\frac{b^2}{4} \sPsi\widetilde\sPsi \p\phi \bar\p \phi+\cdots 
\Big]~.
\eeqn
 Let us   now 
  compute the one-loop correction  to the boundary   two-point function for $\phi$, i.e. 
   $\EV{\sfPhi\sfPhi}$.  It receives  contributions from  several  bubble diagrams (with  the cubic vertices from  \rf{C3}) and a 
self-contraction diagram (with the quartic vertex from \rf{C3}). 
 
 
 There are  two bubble diagrams  where  both  cubic vertices are of the same type:  
    \be
    \la{C4}
 \begin{tikzpicture}[line width=1 pt, scale=1.0, rotate=0,baseline=-0.1cm,decoration={markings, mark=at position 0.53 with {\arrow{>}}}]
\draw (-1.8,0)--(-0.8,0);  
\draw[postaction={decorate}]  (0.8,0) arc(0:180:0.8);
\draw[postaction={decorate}]  (-0.8,0) arc( 180:360:0.8);
\draw  (1.8,0)--(0.8,0); 
\node[above] at (-1.8, 0) {$\sft_1$};
\node[above] at ( 1.8,0) {$\sft_2$};
\node[above] at (-1.0, 0) {$w$};
\node[above] at ( 1.0,0) {$w'$};
\node[left] at (-0.7,-0.3) {$\p\phi$};
\node[right] at (-0.8, 0.3) {$\bar\p \widetilde\sPsi$};
\node[right] at (-0.8, -0.3) {$\sPsi$};
\node[right] at ( 0.7,-0.3) {$\p\phi$};
\node[left] at ( 0.8, -0.3) {$\bar\p \widetilde\sPsi$};
\node[left] at ( 0.8,  0.3) {$\sPsi$};
\end{tikzpicture}
+
 \begin{tikzpicture}[line width=1 pt, scale=1.0, rotate=0,baseline=-0.1cm,decoration={markings, mark=at position 0.53 with {\arrow{>}}}]
\draw (-1.8,0)--(-0.8,0);  
\draw[postaction={decorate}]  (0.8,0) arc(0:180:0.8);
\draw[postaction={decorate}]  (-0.8,0) arc( 180:360:0.8);
\draw  (1.8,0)--(0.8,0); 
\node[left] at (-0.7,-0.3) {$\bar\p\phi$};
\node[right] at (-0.8, 0.3) {$ \p  \sPsi$};
\node[right] at (-0.8, -0.3) {$\widetilde\sPsi$};
\node[right] at ( 0.7,-0.3) {$\bar\p\phi$};
\node[left] at ( 0.8, -0.3) {$ \p  \sPsi$};
\node[left] at ( 0.8,  0.3) {$\widetilde\sPsi$};
\end{tikzpicture}~.
\ee
As these two diagrams are complex conjugate of each other,  it is enough  to  focus on the  contribution of the first one:  
\beqn
I(\sft_1,\sft_2)&=&
\int \rmd^2 w \rmd^2 w' \; \p_{w} g^b_{\phi\phi} (\sft_1,w)\p_{w'} g^b_{\phi\phi} (\sft_2,w')
\bar\p_{w'} g _{\sPsi\widetilde\sPsi} (w,w') \bar \p_{w} g _{\sPsi\widetilde\sPsi} ( w',w)\no
\\&=&
-\int \rmd^2 w \rmd^2 w' \; 
\frac{1}{(\sft_1-w)^2  (\sft_2-w')^2       }    \bar\p_{w'} g _{\sPsi\widetilde\sPsi} (w,w') \bar \p_{w} g _{\sPsi\widetilde\sPsi} ( w',w)~.\la{c5}
\eeqn
Here we used   the notation in \rf{A.4},\rf{2.9}--\rf{2.12} (the free propagators of $\chi,\widetilde \chi$ fields are the same as of 
$\psi, \tilde \psi$). As 
the  integrand is  a rational function  one   may  apply   the residue theorem to do the $\sft,\sft'$ integral.
 It turns out  that no pole   survives,\footnote{
\ This is true for $\sfz>\sfz'$ and $\sfz<\sfz'$,  respectively (recall that $w=\sft+i\sfz, w'=\sft'+i\sfz'$). In the case of $\sfz=\sfz'$, one would encounter a factor of $1/(\sft-\sft')^2$ in the integrand, which leads to a divergence  when performing the $\sft,\sft'$ integral. 
A more careful treatment with an explicit regularization may   lead 
 to  a non-vanishing contribution, but we will not explore this  here.  
 Note that a similar subtlety  happens  also when $\sfz=0$ or $\sfz'=0$.
 }
 implying that the  $\sft,\sft'$ integral gives zero.  Thus   \eqref{C4}   gives a    vanishing contribution.

  The  remaining  bubble diagram   with  two different cubic vertices   and the 
  self-contraction diagram   are  represented  by 
 \be
 \la{C7}
  \begin{tikzpicture}[line width=1 pt, scale=1.0, rotate=0,baseline=-0.1cm,decoration={markings, mark=at position 0.53 with {\arrow{>}}}]
\draw (-1.8,0)--(-0.8,0);  
\draw[postaction={decorate}]  (0.8,0) arc(0:180:0.8);
\draw[postaction={decorate}]  (-0.8,0) arc( 180:360:0.8);
\draw  (1.8,0)--(0.8,0); 
\node[above] at (-1.8, 0) {$\sft_1$};
\node[above] at ( 1.8,0) {$\sft_2$};
\node[above] at (-1.0, 0) {$w$};
\node[above] at ( 1.0,0) {$w'$};
\node[left] at (-0.7,-0.3) {$\bar\p\phi$};
\node[right] at (-0.8, 0.3) {$  \widetilde\sPsi$};
\node[right] at (-0.8, -0.3) {$\p\sPsi$};
\node[right] at ( 0.7,-0.3) {$\p\phi$};
\node[left] at ( 0.8, -0.3) {$\bar\p \widetilde\sPsi$};
\node[left] at ( 0.8,  0.3) {$\sPsi$};
\end{tikzpicture}
+ 
 \begin{tikzpicture}[line width=1 pt, scale=1, rotate=0,baseline=-0.1cm,decoration={markings, mark=at position 0.53 with {\arrow{>}}}]
\draw (-1.8,0)--(1.8,0);  
\draw [postaction={decorate}]  (0.6,0.6) arc(0:180:0.6);
\draw (-0.6,0.6) arc( 180:360:0.6);
\node[left] at (-0.2,-0.3) {$\bar \p\phi$};
\node[right] at ( 0.2,-0.3) {$ \p\phi$};
\node[left] at (-0.5,0.3) {$  \widetilde\sPsi$};
\node[right] at ( 0.5,0.3) {$  \sPsi$};
\node[above] at (-1.8, 0) {$\sft_1$};
\node[above] at ( 1.8,0) {$\sft_2$};
\end{tikzpicture}
\equiv  \Big(\frac{b  }{2 } \Big)^2 \JJ(\sft_1,\sft_2) \ , 
 \ee
 plus  complex conjugate diagrams.  Explicitly, 
 \beqn
\la{C8}
\JJ(\sft_1,\sft_2) &=&
\int \rmd^2 w \rmd^2 w' \; \bar \p_{w} g^b_{\phi\phi} (\sft_1,w)  \p_{w '} g^b_{\phi\phi} (\sft_2,w')
  \p_{w} \bar\p_{w'} g _{\sPsi\widetilde\sPsi} (w,w')  g _{\sPsi\widetilde\sPsi} ( w',w)
 \notag \\&&-
  \int \rmd^2 w  \; \bar \p_{w} g^b_{\phi\phi} (\sft_1,w)  \p_{w } g^b_{\phi\phi} (\sft_2,w')  g _{\sPsi\widetilde\sPsi} ( w,w)
\notag\\&=&
\int \rmd^2 w \rmd^2 w' \; 
\frac{1}{(\sft_1- \bar w)^2  (\sft_2-w')^2       }   \frac{1}{(w-\bar w')^2}   g _{\sPsi\widetilde\sPsi} ( w',w) \ ,
\eeqn
where
the  contribution of the 
self-contraction diagram is exactly cancelled by the part of the bubble diagram associated with the 
    $\delta$-function piece in the derivatives of the propagator (cf. \eqref{2.11}). 
     It is easy to see  that,
\eqref{4.3},\rf{4.6}  and \eqref{C8} happen to differ by an overall factor  only,
although coming from different two-point functions, i.e.
\be
\JJ(\sft_1,\sft_2) =\frac{2}{\pi^2} \widehat{E}(\bt_{12}) =\frac{1}{\sft_{12}^2} \ . 
\ee
Thus    the final  expression for one-loop correction  is given by (taking into account the contribution of the complex conjugate to \rf{C7}) 
\be
\la{C12}
\EV{\sfPhi(\sft_1)\sfPhi(\sft_2)}_{\text{1-loop}}=
2\times \frac {b^2}{4} \JJ (\sft_1,\sft_2) 
=\frac{b^2}{2\sft_{12}^2} \ . 
\ee
Curiously,
  this  is  different from the vanishing result  in  \rf{4.15}. 

Assuming   the symmetry relation  \eqref{B.7},     the result in \rf{C12}  corresponds to 
\be
\la{C13}
\EV{\sfPsi(\sft_1)\tilde \sfPsi(\sft_2)}_{\text{1-loop}} =\frac{b^2}{ \sft_{12}^2}~.
\ee 
This matches  the expression in   
\eqref{4.10}   provided one chooses   $g_0=g(w,w)=\frac12$  (instead of $g_0=1$ in \rf{413}).


One  may  also  compute  the  one-loop correction to the three-point function \rf{4.25}.  Using  \eqref{4.24} and  \rf{C12},\rf{C13}   we get 
\be\la{C10}
 \langle\sfPsi(\sft_1) \widetilde{\sfPsi} (\sft_2)\sfPhi(\sft_3) \rangle^{\text{self-energy}}_{\text{1-loop}}
= -  \frac{3 i\, b^3    }{\sft_{12}\sft_{23}\sft_{13}} \ , 
\qquad 
\langle\sfPsi(\sft_1) \widetilde{\sfPsi} (\sft_2)\sfPhi(\sft_3) \rangle^{\text{triangle}}_{\text{1-loop}}
=  \frac{2  i\, b^3   }{\sft_{12}\sft_{23}\sft_{13}} \big(g_0-1\big)  \ .
\ee
Then  instead of the vanishing  result  in   \rf{4.25} for $g_0=1$  found in section~\ref{sec4}    here we get   
     \be\la{c11} 
          \EV{\sfPhi(\sft_1)\sfPsi(\sft_2)\tilde\sfPsi (\sft_3)}_{\text{1-loop}}  
                 =       \frac{ 
                 ib^3  }{\sft_{12}\sft_{23}\sft_{13}} \big(2 g_0-5 \big)~.
 \ee
 The resulting one-loop   corrected expressions   for the boundary correlators could be, in principle, reconciled with the  corresponding 
 correlators  of the currents provided  the relations between $\kappa$ and $k$  in \rf{2.39} 
 and  between $b$   and $k$ in the action \rf{2.3}  are modified   from their   tree-level   form. 
 
 
 A  more consistent approach  should be 
  to define the boundary correlator with  the ``wave-function" renormalization factors included 
 and that should ensure the invariance of the result   under field  redefinitions. 
 Then the  expressions 
  in  this  Appendix    found starting   with  the redefined  action \rf{C3} 
 could   be reconciled  with the approach  used in section~\ref{sec4}.\foot{\ One  may need also to 
  carefully  take   into  account   contributions 
 of  boundary terms  from   integration by parts. 
  Note also that the use  of the  symmetry relation \eqref{B.7}
 probably   requires a particular choice of the  scheme,  
 i.e.  the value of  the propagator at coinciding  points  $g(w,w)$ and its derivatives.
  For example, using  the  action \eqref{C3} to   compute explicitly the 1-loop correction of the two-point function $\EV{\sfPsi \widetilde \sfPsi  }$, one would encounter the self-contraction  diagram like the third diagram in \eqref{4.1}. The quartic vertex in \eqref{C3} requires us to deal with $\p_w \p_{\bar w} g(w,w)$ due to $\phi$ running in the loop. The regularization of  such 
   derivative term   $\p_w \p_{\bar w} g(w,w)$ was discussed in   \cite{Beccaria:2019dws}. 
  }
  This remains to be clarified further.

\iffa We would like to establish the map between boundary fields and currents following \eqref{2.39}. To that aim,  we find
\be
k\kappa^2 =  \frac{4+b^2}{2}+\mathcal O(b^4),\qquad -2 \sqrt 2 ik \kappa^3 =    -4ib  +  ib^3 (2 g(w,w)-5) +\mathcal O(b^5).
\ee
The solution is 
  \be
 \kappa=\sqrt{\frac{2}{k -1}}, \qquad k+\delta k =\frac{4}{b^2}, \qquad  g(w,w)=\frac{7-\delta k}{4}.
 \ee 
 \fi

%

\bibliography{BT-Biblio}
\bibliographystyle{JHEP}
\end{document}